\documentclass[sigconf]{acmart}

\usepackage{booktabs} 
\usepackage{algorithm}
\usepackage[noend]{algpseudocode}
\usepackage{graphicx}
\usepackage{subcaption}
\usepackage{enumitem}

\newcommand{\Baseline}{Synth }
\newcommand{\sage}{SAGE }
\newcommand{\sagesmall}{sage }
\newcommand{\sagefull}{Synthetic Attenuated Generation Engine }
\newcommand{\siege}{SIEGE }
\newcommand{\siegesmall}{siege }
\newcommand{\siegefull}{Synthetic Isthmus based Edge Generator Engine }

\setcopyright{none}

\acmDOI{}

\acmISBN{}

\acmYear{}
\copyrightyear{}

\acmPrice{}

\begin{document}
\title{New methods to generate massive synthetic networks}

\author{Malay Chakrabarti}
\affiliation{%
  \institution{Virginia Tech}
  \streetaddress{Arlington, VA 22203}
}
\email{malayc@vt.edu}

\author{Lenwood Heath}
\affiliation{%
  \institution{Virginia Tech}
  \streetaddress{Blacksburg, VA 24061}
}
\email{heath@vt.edu}

\author{Naren Ramakrishnan}
\affiliation{%
  \institution{Virginia Tech}
  \streetaddress{Arlington, VA 22203}
}
\email{naren@cs.vt.edu}

\renewcommand{\shortauthors}{M. Chakrabarti et. al.}

\begin{abstract}
One of the biggest needs in network science research is access to
large realistic datasets. As data analytics methods permeate a range of
diverse disciplines---e.g., computational epidemiology,
sustainability, social media analytics, biology, and transportation---
network datasets that can exhibit characteristics encountered in each of these
disciplines becomes paramount.
The key technical issue is
to be able to generate synthetic topologies with pre-specified, arbitrary, 
degree distributions. Existing methods are limited in their ability to
faithfully reproduce macro-level characteristics of networks while at the
same time respecting particular degree distributions. We present a suite
of three algorithms that exploit the principle of residual degree attenuation
to generate synthetic topologies that adhere to macro-level real-world
characteristics. By evaluating these algorithms w.r.t.
several real-world datasets we demonstrate their ability to faithfully
reproduce network characteristics such as node degree,
clustering coefficient, hop length, and k-core structure
distributions.
\end{abstract}

\maketitle

\section{Introduction}
Network science has made inroads into a variety of domains ~\cite{kleinberg1999web},

\noindent
~\cite{kumar2000stochastic}, ~\cite{white1986structure}, ~\cite{camacho2002robust}, ~\cite{guelzim2002topological}, ~\cite{jeong2003measuring}, ~\cite{ebel2002scale} including
epidemiology, sustainability, health informatics, and transportation.
As the scope of such applications continues to broaden, data analytics
research needs a constant supply of realistic datasets that
mimic characteristics encountered in these domains.

\begin{figure}[h]
\centering
\includegraphics[scale=0.45]{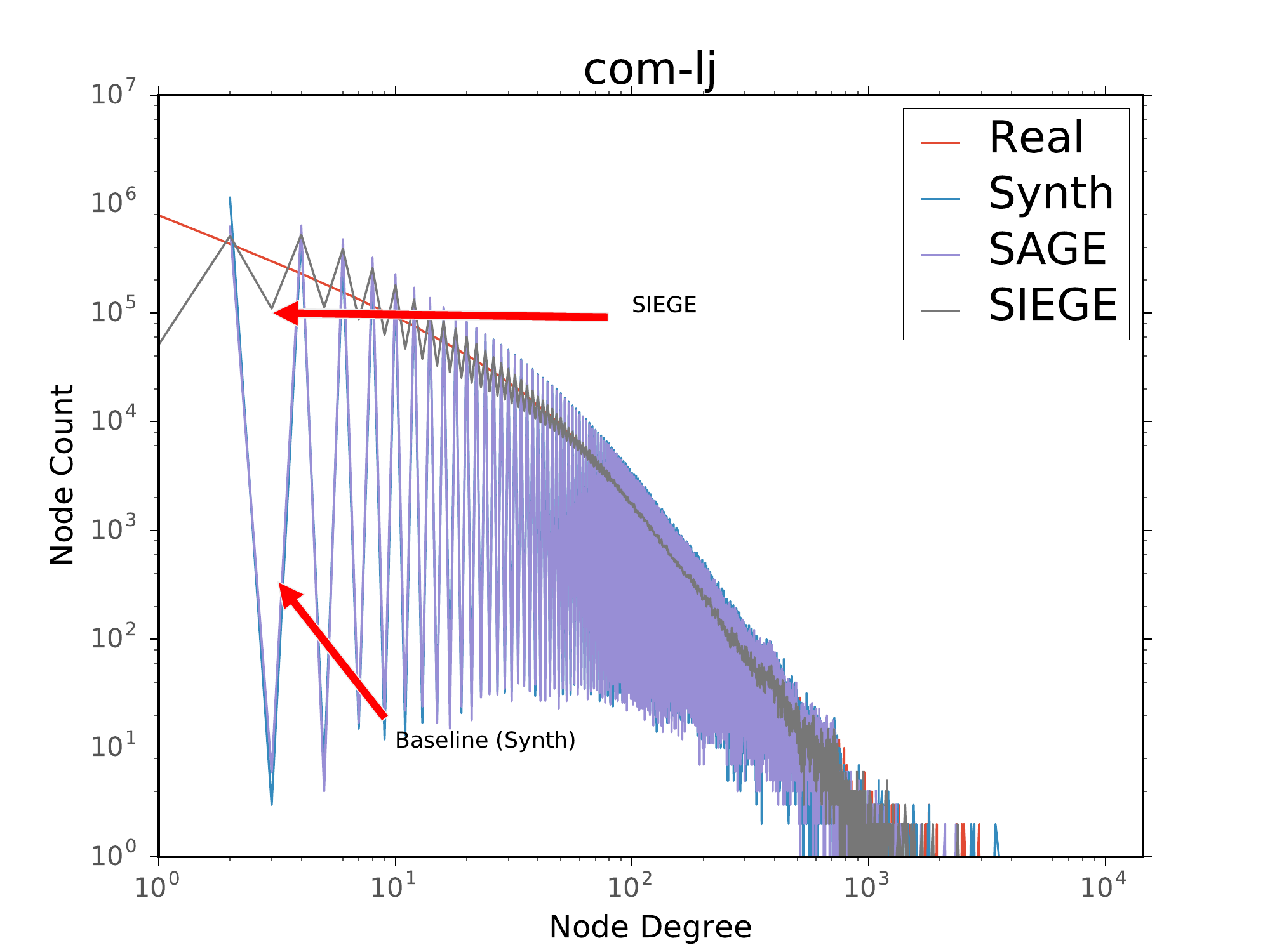}
\caption{This plot demonstrates how our proposed model \textbf{\siege} captures real topology characteristics better than the baseline \textbf{\Baseline} when
used to mimic the com-LiveJournal dataset (3.99 million nodes and over 
34 million edges).}
\end{figure}
\noindent   
A key need is to have synthetic network representations that reproduce
macro-level behavior found in real networks. Examples of phenomena
we wish to study using synthetic networks include:
\textbf{robustness:-}  against targeted and random removal of nodes and edges, \textbf{homophily:-} whether nodes that share similar characteristics tend to have more edges between them, \textbf{search:-} to understand the impact on search strategy with respect to network topology, and 
\textbf{diffusion processes:-} the spread of dynamics over a network.
Having realistic synthetic network representations supports privacy concerns
and can help realize what-if studies that are not possible to conduct
in real life. We abstract the essential problem as one of creating synthetic
networks with arbitrary, user-specified, degree distributions. 

The primary contributions of this paper are:

\begin{itemize}[leftmargin = 0.4cm]
\item \textbf{Weakness analysis.} We begin by demonstrating that
state-of-the-art techniques for generating networks with
arbitrary degree distributions are found wanting. In particular, we
demonstrate that the well known DEG method~\cite{heath2011generating} falls short when
evaluated against real characteristics 
(broadly summarized as triadic closures growing superlinearly in the 
number of edges in the topology).

\item \textbf{Proposal of new models.} We propose a fail safe baseline \Baseline that is more stable than DEG, and propose two more models \sage, \siege that outperform it, in terms of generating synthetic topologies with adherence to macro level real world characteristics. This can be seen using the annotations in the \textbf{Figure 1 (shown in log-log scale)}. \sage and \siege uses the principle of residual degree attenuation (to be described in a later
section).

\item \textbf{Extensive experiments and evaluations to validate the effectiveness of our proposed models.} We compare our synthetically generated networks
against 25 real-world datasets w.r.t. five 
well accepted characteristics, demonstrating consistent superiority in
reproducing these characteristics.
\end{itemize}

\section{Related Work}
Related research
in this area can be understood from three viewpoints \cite{nettleton2015generating}: synthetic topology generation without data, synthetic topology generation followed by data generation to fit the topology  (\cite{barrett2009generation}, \cite{perez2015synthetic}) and finally the notion of like liking like otherwise known as homophily (and it's counterpart concept diversity) \cite{mcpherson2001birds}.
Among these three viewpoints the first has the lions share in terms of 
past research. Note that we will touch upon the 
related work from this viewpoint, as our work serves to extend this approach.
  
\subsection{Synthetic topology generation without data}

\begin{figure}[t]
\centering
  \begin{subfigure}[b]{0.60 \linewidth}
    \includegraphics[width=.99\textwidth]{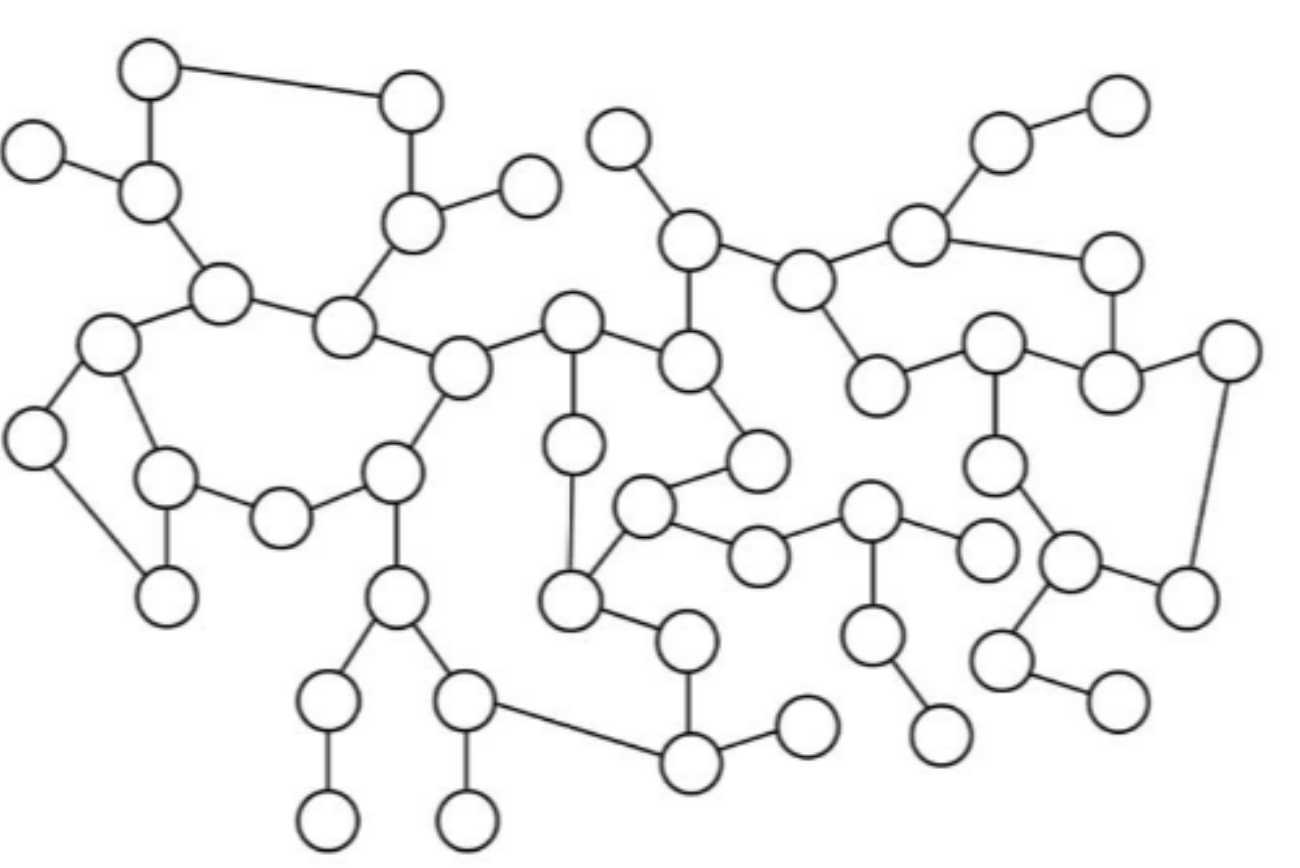}
    \caption{Random Topology}\label{fig:1a}
  \end{subfigure}%

  \begin{subfigure}[b]{0.60 \linewidth}
    \includegraphics[width=.99\textwidth]{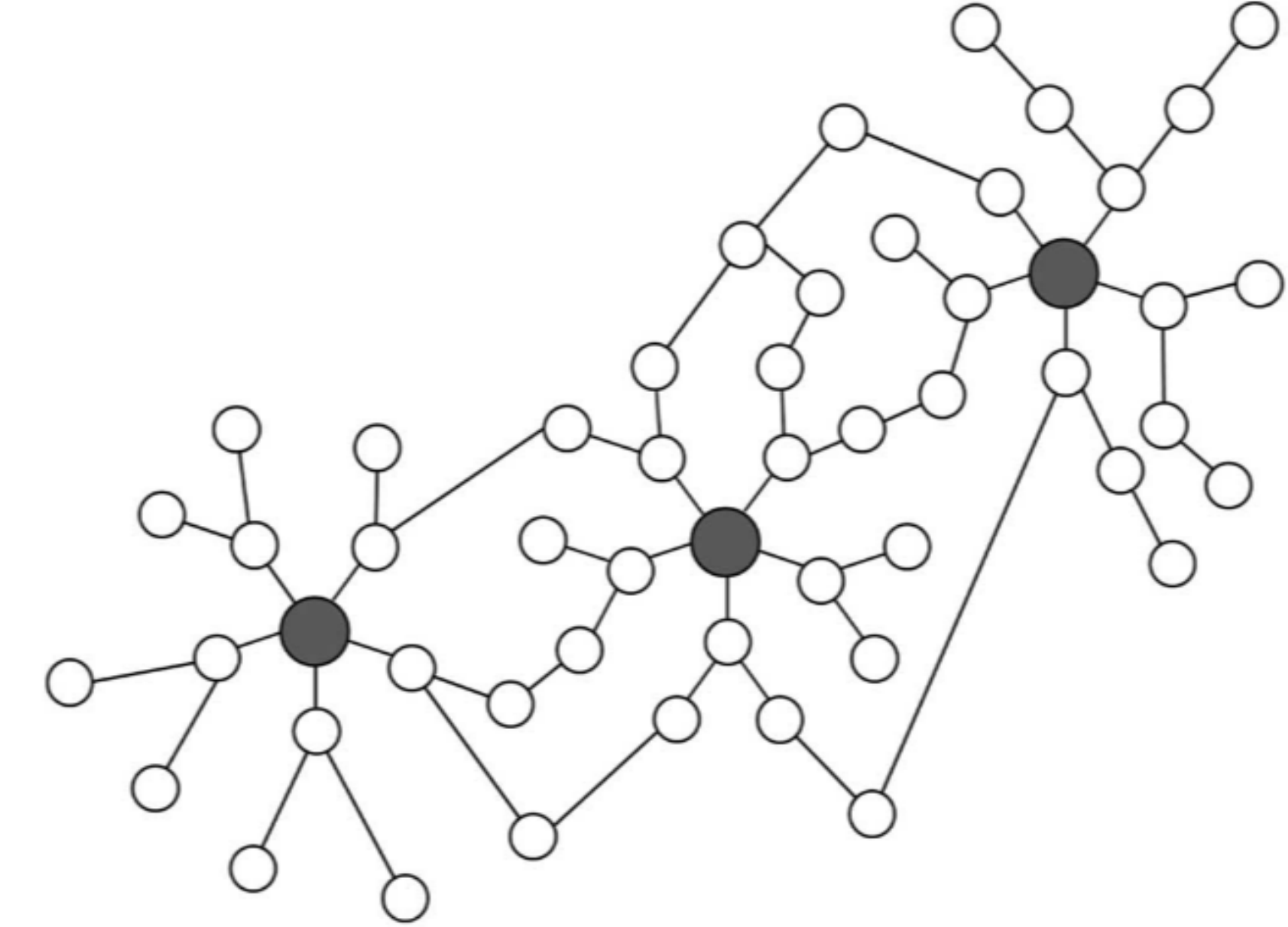}
    \caption{Scale-Free Topology}\label{fig:1b}
  \end{subfigure}%
  \caption{A random topology model such as the Erdos-Renyi model
converges to a Poisson degree distribution while a scale-free topology
tends to have fewer nodes with high degrees and more nodes with 
smaller degrees. On a log-log scale of node count vs node degree the curve would be linear, oblique and heavy tailed.}
\end{figure}

\subsubsection{ER Model:} 
Solomonoff et al.\cite{solomonoff1951connectivity} and Erdos et al.\cite{erdos1960evolution} independently presented a model of graph topology generation with n nodes where each pair of nodes can  have an edge between them with a probability p. 
In this method the probability of having a graph with n nodes and l edges is ${L \choose l} p^l(1-p)^{L - l}$ , where $L = {n \choose 2}$, which is the maximum possible number of edges in a graph. 
The degree distribution is approximately Poisson. 
Networks generated by this model do not satisfy properties such as the
small-world phenomenon.

\subsubsection{Small-world  Networks:}
Networks generated by the ER model do not exhibit local clustering and triadic closures. The degree distribution converges to a Poisson distribution
instead of a power law which most scale-free real world networks tend to exhibit (see Fig.~2).
Watts et al.\cite{watts1998collective} proposed a small-world
model, for n nodes, mean degree k and parameter p where $0 \leq p \leq 1$, to produce a network with n nodes and nk/2 edges. The limitation of this method is that it generates non-scale free networks.
The two step generation process works as follows:

1) Construct a regular ring lattice with n nodes each connected to k of its neighbors with k/2 nodes on each side.

2) Rewire every edge with probability p,  such that $(n_i, n_j)$ is replaced with $(n_i, n_k)$ where k is chosen with uniform probability from all possible values such that there are no self-loops and link-duplication.

The degree distribution of the topology is a Dirac-delta function centered at k for the ring lattice and is Poisson distributed in the limiting case of $p \to 1$. If $p=0$ then the graph is a regular ring lattice with clustering coefficient $C = \frac{3k -3}{4k - 2}$ and the diameter $\frac{n}{4k}$ for large n. If $p=1$ then the graph is completely random with diameter $\frac{\log n}{logk}$ and clustering coefficient $C = \frac{2k}{n} $ which is a low value. Between the two extremities of $p = 0 $ and $ p = 1$ there exists a region that gives low diameter and high clustering coefficient, the charactristics of a small-world
network.

To overcome its inability to generate scale-free networks,
Newman et al.\cite{newman1999renormalization} proposed a variant where for each edge in the network with probability $p$ a new edge is added between two randomly chosen nodes. The degree distribution of this network is also non-conforming to real world as it is uniform for the lattice and binomially 
distributed for the randomly added edges.

\subsubsection{Preferential Attachment Model.}

This class of algorithm is based on the principle of "cumulative advantage" as stated by Price \cite{price1976general} to describe citation networks: 
the probability that one comes across a paper while reading the
literature is intuitively proportional to the number of papers citing it.
Barabasi et al.\cite{barabasi1999emergence} rediscovered the notion by considering an undirected graph and named the phenomenon as `preferential attachment.' 
In their algorithm the network initially begins with $m_0$ connected nodes. New nodes are added to the existing network one at a time. Each new node is connected to m existing nodes in the network where $m \leq m_0$ and the probability $P$ that a new node connects to an existing node $v$ is proportional to the degree of $v$, $d[v]$. Thus $P(d[v]) = \frac{d[v]}{\sum_{i=1}^{n} d[i]}$.
Barabasi et al.~\cite{barabasi1999emergence} also showed that this method leads to topologies with degree distribution that follows a power law $p_k \propto k^{- \alpha}$ where $\alpha = 3$. The clustering coefficient also increases with network size as $C \approx N^{-0.75}$, a behavior that is different from small-world networks where clustering tends to be independent of system size. 
Adamic et al.~\cite{adamic2000comment} 
provided empirically observed properties of the web link structure 
to highlight shortcomings of the Barabasi et al. model.
Several variants of the PA model exist, a selected list of which is listed here. Holme et al. \cite{holme2002growing} proposed a model that at first has a preferential attachment step where a new node v is attached to an existing node w with the probability proportional to the degree of w. This is followed by a triad formation step.  where the new node v is connected to randomly chosen neighbor u of the node w that was selected as attachment site in the PA step.  If there remains no pair to be 
connected, i.e., if all neighbors of w are already connected to
v, then a PA step is done instead.
Guo et al.~\cite{guo2006growing}, proposed a simple algorithm that generated a scale-free small world network that is similar to Holme et al.~\cite{holme2002growing} in that it too adds a new node v with m edges at each time step. 
In the first step, each edge of v is attached to an existing node with a probability proportional to it's degree. The second step involves attaching the new node v to a randomly chosen neighbor s of w (w being the nodes which has edges between itself and v) with probability given by 
$ p_s = \frac{d[s]^{\beta}}{\sum_{i \in r_w} d[i]^{\beta}}$ where d[i] is the degree of node i, $r_w$ is the set of neighbors of w and $\beta$ is the preferential exponent parameter. 
This model can generate a non-trivial clustering property.
Wang et al.~\cite{wang2008evolving} introduced a model that at each time step added a new node v with m edges. A node u is selected at random such that the local world of v is u(s) which is the set of all nodes that are at a distance of s or less from u. The node v does a preferential attachment to node $w \in u(s)$. Followed by which v connects itself to the highest degree neighbor of node w with a probability p, which also acts a parameter for tuning clustering coefficient.
Schank et al.~\cite{schank2004approximating} proposed a model where at each time step a new node with m edges is added to previously existing nodes by following a PA step. Then as a second step, for each node $v \in V$, two neighbors u and w are selected at random and are connected. The clustering coefficient can be varied by varying the number of times the second step is repeated.

Guo et al.\cite{guo2009random} proposed a model that generated  directed graphs where in-degree and out-degree follows power law.      
The algorithm uses the power exponent for in-degree and out-degree to generate the in-degree and out-degree sequence along with number of nodes n and edges m. 
Following this step,
out-stubs and in-stubs are randomly assigned to vertices as per the degree sequence. Finally, links are created between randomly between in-stubs and out-stubs with some probability. The average clustering coefficient ($C_f$) of the graph thus obtained tends to be lower than the specified input. 
In those cases edge switching occurs to increase the $C_f$ for some nodes, according to the following set of rules:- A random node $v$ and four distinct nodes $w,x,y,$ and $z$ are chosen, such that, 

1. $w$ and $x$ are neighbors of $v$

2. $y$ and $z$ are not neighbors of $v$

3. $(w,x) \notin E$ and $(z,y) \notin E$

4. $(w,y) \in E$ and $(z,x) \in E$

5. $C(v)$ of neighbors shared by $w$ and $y < target C_f$

6. $C(v)$ of neighbors shared by $x$ and $z < target C_f$

\noindent
On deleting edges $(w,y)$ and $(z,x)$ from $E$, and adding edges $(w,x)$ and $(z,y)$ to E would increase the local $C_f$ of the node $v$, which contributes to increasing global $C_f$ of the network.
Gleeson et al.\cite{gleeson2009bond} proposed a model that uses the joint probability distribution function $\gamma(k,c)$, which gives the probability that a randomly chosen node in the network will have a degree k and would belong to a c-clique. The network is generated by connecting stubs (viewed as external edges) to these cliques. This method generates graphs with n-adic structures as opposed to just triadic as is prevalent in many methods. 
However, it is still not realistic as real-world networks need 
not have nodes that are constrained to be part of a single clique.

\subsubsection{Configuration Model}

Molloy et al.~\cite{molloy1995critical} proposed the configuration model, wherein given the degree distribution (d[1], d[2] ... d[n]) of n nodes as input it creates a set Q having d[i] distinct copies of each node i and then selects two nodes at random from the set Q and adds an edge between them. This process is repeated until every element is part of some edge.
It produces a multigraph from which a simple graph can be obtained by merging the multi-edges and removing the self-loops.
Heath et al.~\cite{heath2011generating} proposed a model called DEG that 
takes a degree sequence and a target clustering coefficient and tries to generate a simple random graph with these properties. It is based on the configuration model, i.e.,
each node $v$ is seen as having a finite number of stubs coming out of it, which is equal to the degree of the node $d[v]$. The next step, in this model, is to calculate the total number of triangles that are required to achieve the given clustering coefficient, and then create those by selecting three stubs, guided by the residual degree distribution, and joining them to create a triangle. Once the estimated number of triangles is created, a pair of stubs are selected, again guided by the residual degree distribution, and joined to create additional edges.
This model is limited in generating synthetic networks that conforms to real world network statistics if the total number of triangles required exceeds the total number of edges provided in the graph, which almost invariably is the case with real world networks as they follow power law degree distribution. To overcome this shortcoming, we propose three new DEG-inspired models and measure the statistics they generate with the real world network statistics we choose to replicate. 

\section{Definitions}

Let $G =(V,E)$ be an undirected simple graph with a set of nodes $V$ and a set of edges $E$. The cardinality of $V$, i.e $|V| = n$ and $|E| = m$. The degree $d(v) = |{u \in V : \exists \{u,v\} \in E}|$ of a node $v$ is the count of all nodes that are adjacent to $v$. A complete graph on n nodes is $K_{n}$, and a complete r-partite graph with partition of size $(n_1, n_2...n_r)$ is $K_{n_{1},...n_{r}}$. A triangle $\Delta = (V_{\Delta}, E_{\Delta})$ of a graph $G = (V,E)$ is a three node subgraph with $V_{\Delta} = \{u,v,w\} \subseteq V$ and $E_{\Delta} = \{(u,v),(v,w),(u,w)\} \subseteq E.$ The number of triangles at a node $v$ is given by 
\begin{equation}
\Delta(v) = |\{(u,w) \in E | (v,u) \in E \ and \ (v,w) \in E\}|.
\end{equation}
The number of triples at a node v is the number of length 2 paths in which v is the central node. It can be quantified as
\begin{equation}
\tau(v) = {d[v] \choose 2}.
\end{equation}

The local clustering coefficient $C(v)$ is the ratio of the number of triangles at node v to the number of triples,, i.e,

\begin{equation}
C(v) = \frac{\Delta(v)}{\tau(v)} \ \mbox{if} \d[v] \geq 2 .
\end{equation}

The clustering coefficient of the graph $C(G)$ for the graph $G(V,E)$ is defined as the average of $C(v)$ of all nodes $v$ in the graph. That is,

\begin{equation}
C(G) = \frac{1}{|V|} \sum_{v \in V}C(v).
\end{equation} 
 
The total number of triangles in a graph in terms of $\Delta(v)$ is 
\begin{equation}
\Delta(G) = \frac{\sum_{v \in V} \Delta(v)}{3}
\end{equation}
as each triangle constitutes three nodes.

The number of triples in the graph is 
\begin{equation}
\tau(G) = \sum_{v \in V}\tau(v). 
\end{equation}  

The global clustering coefficient $C_g(G)$ is defined as three times the number of triangles in the graph to the number of triples in the graph, that is

\begin{equation}
C_g(G) = \frac{3\Delta(G)}{\tau(G)}
\end{equation}

If the total number of connected components in the graph is $\lambda$, then a cut edge or edge bridge $(u,v) \in E$ is an edge which if removed from the graph, changes the total number of connected components to $\lambda^{'}$ where $\lambda^{'} - \lambda = 1$.
A path from a node $u$ to $v$ in a graph $G$ is a sequence of edges $\{(x_0,x_1),....,(x_{n-1},x_n)\}$, where $x_0 = u$ and $x_n = v$. The hop length is the one that has the minimum number of edges among all such possible paths between any two given vertices in the graph.  
Finally, the connected components that are left after all vertices of degree less than k have been removed are defined as the k-cores of the graph.

\subsection{Dynamics on Networks}
The above definitions capture structural properties of networks. We are
also interested in dynamics exhibited by networks. For the purpose
of this paper, we focus on the SIR epidemiological model ~\cite{kermack1927contribution} wherein
a node can go through three stages during the course of the dynamics:

\begin{itemize}[leftmargin = 0.4cm]
\item Susceptible: Before the node has been infected it is susceptible to infection from any of its neighbors.

\item Infectious: Once infected, it is capable of infecting any of its susceptible neighbors 

\item Recovered: Once the required time period of infection for that node is over it is considered as recovered.

\end{itemize}
  
Assuming S is the proportion of population that is susceptible, I is the proportion of the population that is infected and R is the proportion that has recovered, $\beta$ being the transmission rate among the individuals and $\gamma$ being the recovery rate we have,    

\begin{equation*}
\frac{dS}{dt} = - \beta S I \ \text{,}\ \frac{dI}{dt} =  \beta S I - \gamma I  \ \text{,}\ \frac{dR}{dt} = \beta  I 
\end{equation*}

This also assumes there is no birth, death or migration of populations. Thus new nodes are not allowed to be created. Our goal is to evaluate our synthetic
networks against SIR dynamics imposed on them versus SIR dynamics on real
networks.
 
\section{Models}

\subsection{Making DEG work in practice}
It is noted that an n-clique has exactly $n \choose 3$ triangles and asymptotically $\Delta(G_{n-clique}) \in \Theta(n^3)$. We also know that $|E| \in \Theta(n^2) $. Thus $\Delta(G_{n-clique}) \in \Theta(|E|^{3/2})$ \cite{schank2005finding}. A n-clique has by definition maximum number of edges possible in a n node graph. If we fix the total number of nodes and want to accommodate as many edges as possible within the clique we can have a family of Graphs $G_{|E|}$ which are not cliques but $\Delta(G_{|E|})\ \mbox{would still be} \  \Theta(|E|^{3/2}))$. This shows, it is very much possible that networks can have more triangles than the total number of edges and in all those cases DEG fails.     
We propose an algorithm \textbf{\Baseline}, that gets around this problem. The change proposed is seen in line 7 where the number of triangles to be created is rationed by the number of edges available. For most real world networks, the number of triangles that needs to be created exceeds the number of edges provided. In those cases \textbf{\Baseline} will run to completion as opposed to DEG. However, it will have a higher chance of generating a singular connected component (or few but large connnected components) as a representation of the graph for most real networks as all edges are exhausted in an effort to reach the delineated number of triangles. This  might not be representative of the actual structure which tends to have multiple connected components. It should be noted that it is not a rigorous necessity for the total number of triadic closures, to be more than the total number of edges in order for DEG to stall. Even if it is reasonably close to the number of edges, the algorithm can still stall given the fact that, on an average it consumes more than one unique edge in order to generate a triadic closure. (The oregon010331 to oregon010428 dataset in \textbf{Table 1} 
described later serve well to illustrate the purpose.) The algorithm DEG also re-computes the residual degree distribution after each successful sampling of at-most 3 unique edges. This does not affect the runtime as long as the count of the following two factors are within reasonable  bounds:

\begin{itemize}[leftmargin = 0.4cm]
\item $ |E_{G_n}| << |E_{G_{n-clique}}| $, as this evidently reduces the number of loop iterations.       
 
\item $ |V| \approx 10^3$ , as this results in fast re-normalization of the residual degree distribution once 3 (at-most) unique edges are sampled.
\end{itemize}

But as we attempt to generate larger synthetic datasets 
the aforementioned factors tend not to hold true anymore thus leading to scalability issues. To overcome this issue,
we conduct bulk edge sampling by taking a snapshot of the residual degree distribution at iteration points which are tuned by the step size of our choice. This automatically reduces the number of re-normalization computation by a factor of step-size thus also reducing the computation time by a significant amount while sacrificing some accuracy. However, since this bulk sampling will be applied with uniform step-size for any given large dataset across all the algorithms proposed, the relative accuracy among them should not be affected. It should be noted that this does not mean that the
three algorithms will have the same absolute error margins, but that the relative error variations should be invariant given a non-zero step size. 

\subsection{New Models}





\begin{algorithm}[ht]
\caption{\textbf{\Baseline}}\label{SYNTH}
\begin{algorithmic}[1]
\Procedure{$\Baseline$}{n,d,$C_{g}(G)$,Step}
\State $V \gets \{1,2 ... ,n\}$ 
\State $E \gets \phi$ 

\State $M \gets \frac{\sum_{x \in V}d[x]}{2}$ 

\State $rd \gets d$ 

\State $T \gets \frac{C_{g}(G) \sum_{i = 1}^{n} {d[i] \choose 2}}{3}$ 
\While {$T > 0 \ \textit{and} \ M > 0$}
	\State $\textit{smpl} \gets 0$	
	\State $\mbox{Compute} \forall p \in V \frac{rd[p]}{\sum_{x \in V} rd[x]}$
	\While {$T > 0 \ \textit{and} \ M > 0 \textit{and smpl} < Step$}
		\State $\mbox{Choose three distinct nodes}$
		\State $\mbox{u with probability} \frac{rd[u]}{\sum_{x \in V} rd[x]}$
	
		\State $\mbox{v with probability} \frac{rd[v]}{\sum_{x \in V} rd[x]}$
	
		\State $\mbox{w with probability} \frac{rd[w]}{\sum_{x \in V} rd[x]}$
		\State $\mbox{Check} \ rd[u], rd[v], \mbox{and} \  rd[w]$
		\State $\mbox{Add edges needed to create a triangle}$ 	\State $\mbox{between the chosen nodes} \  u,v, \mbox{and} \ w$
\State $\mbox{update} \ rd[u], rd[v], rd[w], \ T,\textit{smpl} \ \mbox{and} \ M \ \mbox{accordingly}$
	\EndWhile
\EndWhile
\While {$M > 0$}
	\State $\textit{smpl} \gets 0$	
	\State $\mbox{Compute} \forall p \in V \frac{rd[p]}{\sum_{x \in V} rd[x]}$
	\While {$M > 0 \textit{and smpl} < Step$}
		\State $\mbox{Choose two distinct nodes}$
		\State $\mbox{u with probability} \frac{rd[u]}{\sum_{x \in V} rd[x]}$
		\State $\mbox{v with probability} \frac{rd[v]}{\sum_{x \in V} rd[x]}$
		\State $smpl \gets smpl + 1$
		\If {$(u,v) \notin E$}
			\State $E \gets E \cup \{(u,v)\}$
			\State $rd[u] \gets rd[u] - 1$
			\State $rd[v] \gets rd[v] - 1$
			\State $M \gets M - 1$
		\EndIf
	\EndWhile	
\EndWhile
\Return{$(V,E)$}
\EndProcedure
\end{algorithmic}
\end{algorithm}




	
	
	
	

\subsubsection{\sagefull(\sage)}

We propose another variant, \textbf{\sage} where the residual degree of $u,v$ is reduced by $1$ if a new edge $\{(u,v)\}$ is added to the graph but the decrement is attenuated after the residual degree hits 1. The edge addition from line 9 onwards, however, allows the node degree to reduce to 0, as that ensures a node that contributes a new edge is not selected again. Note this holds good even in the worst case when there is no triangle in the graph.In such a scenario the maximum possible number of edges $m$ to fill would be $\lfloor \frac{n^2}{4} \rfloor$  (Mantel's Theorem) \cite{mantel1907problem} which is only possible on a complete bipartite graph $K_{\lfloor \frac{n}{2} \rfloor, \lceil \frac{n}{2} \rceil}$, and can be successfully processed with a uniform degree distribution, which is almost always the outcome after line 8, for all executions of generating synthetic topology imitating real world networks. The rationale for keeping step size has already been covered in the previous section.

\begin{algorithm}[ht]
\caption{\textbf{\sage}}\label{SYNTH1}
\begin{algorithmic}[1]
\Procedure{$\sage$}{n,d,$C_{g}(G)$,Step}
\State $V \gets \{1,2 ... ,n\}$ 
\State $E \gets \phi$ 

\State $M \gets \frac{\sum_{x \in V}d[x]}{2}$ 

\State $rd \gets d$ 

\State $T \gets \frac{C_{g}(G) \sum_{i = 1}^{n} {d[i] \choose 2}}{3}$

\State Repeat line 7 to 18 from \Baseline 
\State but $\mbox{update} \ rd[u], rd[v], rd[w],\  \mbox{only if they are} > 1$
\While {$M > 0$}
	\State $\textit{smpl} \gets 0$	
	\State $\mbox{Compute} \forall p \in V \frac{rd[p]}{\sum_{x \in V} rd[x]}$
	\While {$M > 0 \textit{and smpl} < Step$}
		\State $\mbox{Choose two distinct nodes}$
		\State $\mbox{u with probability} \frac{rd[u]}{\sum_{x \in V} rd[x]}$
	
		\State $\mbox{v with probability} \frac{rd[v]}{\sum_{x \in V} rd[x]}$
		\State $smpl \gets smpl + 1$
		\If {$(u,v) \notin E$}
			\State $E \gets E \cup \{(u,v)\}$
			\If {$rd[u] \geq 1$}
				\State $rd[u] \gets rd[u] - 1$
			\EndIf
			\If {$rd[v] \geq 1$}
				\State $rd[v] \gets rd[v] - 1$
			\EndIf
			\State $M \gets M - 1$
		\EndIf	
	\EndWhile
\EndWhile
\Return{$(V,E)$}
\EndProcedure
\end{algorithmic}
\end{algorithm}

\subsubsection{\siegefull (\siege)}

Finally, we propose another variant \textbf{\siege} that uses the total number of edge bridges (isthmus) in the original network, as well as the target clustering coefficient to create the synthetic network. \cite{tarjan1974note} proposed a linear time model to find the isthmus' in a given graph. We can modify that a little by keeping a counter (captured by the input parameter \textbf{$EB_{Count}$} (edge bridge count, please refer table 2), which we provide as input to \textbf{SIEGE}) and thus find the cardinality of the set of isthmus' in the graph. Also this approach does not require a single edge degree distribution to be stored as compared to \cite{newman2009random} making it space efficient.  In this algorithm, when execution is at line 3, then either of the two cases happen:-

\begin{itemize}[leftmargin = 0.4cm]
\item The total number of triangles required is generated and more edges are available. In this case the residual edge count is replaced by edge bridge count and the execution is purely focused on generating new edges which has a much higher probability of generating stubs as opposed to triangles.

\item In the other scenario, when the edge count is exhausted without reaching the goal of the requisite number of triangles, the algorithm then purely generates single edges from an uniform residual degree distribution, and residual degree count is allowed to reach 0, such that unique nodes can be selected at each iteration.
\end{itemize}

The topologies generated by this on an average comes closest to real world networks as compared to \textbf{\Baseline} and \textbf{\sage}. However, this method uses more edges than the original network topology provides.

\begin{algorithm}[ht]
\caption{\textbf{\siege}}\label{SYNTH2}
\begin{algorithmic}[1]
\Procedure{$\siege$}{n,d,$C_{g}(G)$,Step,$EB_{Count}$}
\State Repeat line 2 to 8 from \sage
\While {$EB_{Count} > 0$}
	\State $\textit{smpl} \gets 0$	
	\State $\mbox{Compute} \forall p \in V \frac{rd[p]}{\sum_{x \in V} rd[x]}$
	\While {$EB_{Count} > 0 \textit{and smpl} < Step$}
		\State $\mbox{Choose two distinct nodes}$
		\State $\mbox{u with probability} \frac{rd[u]}{\sum_{x \in V} rd[x]}$
	
		\State $\mbox{v with probability} \frac{rd[v]}{\sum_{x \in V} rd[x]}$
		\State $smpl \gets smpl + 1$
		\If {$(u,v) \notin E$}
			\State $E \gets E \cup \{(u,v)\}$
			\If {$rd[u] \geq 1$}
				\State $rd[u] \gets rd[u] - 1$
			\EndIf
			\If {$rd[v] \geq 1$}
				\State $rd[v] \gets rd[v] - 1$
			\EndIf
			\State $EB_{Count} \gets EB_{Count} - 1$
		\EndIf	
	\EndWhile
\EndWhile
\Return{$(V,E)$}
\EndProcedure
\end{algorithmic}
\end{algorithm}

\section{Experimental Setup}

\subsection{Datasets}

We downloaded twenty five topology datasets/statistics
(detailed in \textbf{Table 1}) from SNAP (the Stanford Network Analysis Project) and used these statistics as input parameters to our three algorithms to generate three synthetic topologies for each. The twenty five topologies are broadly classified into six categories which are listed as follows:

\begin{table}
\centering
\scriptsize
\caption{It can be seen that the number of triangles are either more or close to the number of edges in the dataset. DEG works primarily if the number of triangles are significantly less than the number of edges. Most real world networks do not adhere to this characteristic.}
\begin{tabular}{ |c||c|c|c|c|c|c|  }
 \hline
 \multicolumn{7}{|c|}{$\textrm{Topology \ Characteristics} $} \\
 \hline
 $\textrm{Topology}$& $|V|$ & $|E|$  & $|\Delta|$&$\textrm{Diam}$ & $\textrm{EDiam}$ & $\textrm{Step}$\\
 \hline 
egoFacebook   & 4039 & 88234  & 1612010 &  8  & 4.7 & 1 \\
caGrQc & 5242 & 14496  & 48260   & 17 & 7.6 & 1\\
caHepTh & 9877 & 25998  & 28339 & 17	 & 7.4 & 1\\
oregon010331  & 10670 &22002 &  17144&  9 & 4.4 & 1\\
oregon010407 & 10729 & 21999  & 15834 & 11 & 4.5 & 1\\
oregon010414 & 10790 &22469  & 18237   & 9 & 4.4 & 1\\
oregon010421 & 10859 &22747 &  19108 & 10 & 4.4 & 1\\
oregon010428 & 10886 &22493 &  17645 & 10 & 4.4 & 1\\
oregon020331 & 10900 &31180 &  82856 & 9 & 4.3 & 1\\
oregon020407 & 10981 &30855 &  78138 & 11 & 4.3 & 1\\
oregon020414 & 11019 &31761 &  88905 & 8 & 4.2 & 1\\
oregon020421 & 11080 &31538 &  82029 & 9 & 4.3 & 1\\
oregon020428 & 11113 &31434 &  78000 & 9 & 4.2 & 1\\
oregon020505 & 11157 &30943 &  72182 & 9 & 4.3 & 1\\
oregon020512 & 11260 &31303 &  72866 & 9 & 4.2 & 1\\
oregon020519 & 11375 &32287 &  83709 & 9 & 4.3 & 1\\
oregon020526 & 11461 &32730 &  89541 & 9 & 4.3 & 1\\
email-Enron & 36692 &183831 &  727044 & 11 & 4.8 & 1000\\
loc-gowalla & 196591 &950327 &  2273138 & 14 & 5.7 & 1000\\
loc-brightkite & 58228 & 214078 &  494728 & 16 & 6 & 1000\\
com-amazon & 334863 &925872 &  667129 & 44 & 15 & 10000\\
com-dblp & 317080 & 1049866 &  2224385 & 21 & 8 & 10000\\
com-youtube & 1134890 & 2987624 &  3056386 & 20 & 6.5 & 20000\\
com-lj & 3997962 & 34681189 &  177820130 & 17 & 6.5 & 40000\\
as-skitter & 1696415 & 11095298 &  28769868 & 25 & 6 & 20000\\
 \hline
\end{tabular}
\end{table}

\begin{itemize}[leftmargin = 0.4cm]
\item \textbf{Social Network:} This domain includes \textbf{ego-Facebook}~\cite{mcauley2012learning}which is an undirected topology dataset collected from survey participants using the facebook app. This dataset is completely anonymized by substituting the facebook internal ids with a new value.

\item \textbf{Autonomous Systems Based Network:} We use fifteen topologies from this domain, five of which are \textbf{ oregon1$\_$010 \{331 to 428\}} \cite{leskovec2005graphs}. These undirected topologies are representations of autonomous systems peering information that were inferred from Oregon route views from $31^{st} \ March \ 2001$ to $28^{th} \ April  \ 2001$. Nine, (\textbf{oregon$\_$020 \{331 to 526\}}), are generated by combining Oregon route-views, looking glass data, and routing registry. Finally we have, \textbf{as-skitter} which is an undirected internet topology graph, from traceroutes run daily in 2005. It has around 1.7 million nodes and 11 million edges.

\item \textbf{Collaboration Network:} The topology \textbf{ca-Gr-Qc} is an Arxiv GR-QC (General Relativity and Quantum Cosmology) collaboration network and covers scientific collaborations between authors papers submitted to the
General Relativity and Quantum Cosmology category. The topology \textbf{ca-Hep-Th} represents the same idea but for the 
High Energy Physics category.

\item \textbf{Communication Network:} The topology \textbf{email-Enron} is an Enron email communication network~\cite{leskovec2009community,klimt2004introducing}. This covers around half million emails and was originally made public by the Federal Energy Regulatory Commission during its investigation. Nodes of the network are email addresses and if an address i sent at least one email 
to address j, the graph contains an undirected edge from i to j. 

\item \textbf{Location based OSN:} The topology \textbf{loc-gowalla} is from Gowalla which is a location-based social networking website where users share their locations by checking-in. This data~\cite{cho2011friendship} was collected using public API from a total of 6,442,890 check-ins of these users over the period of Feb. 2009 - Oct. 2010. We also use another topology \textbf{loc-brightkite} that is similar to the Gowalla based dataset~\cite{cho2011friendship} which was collected from a total of 4,491,143 checkins of these users over the period of Apr. 2008 - Oct. 2010. 

\item \textbf{Networks with ground-truth communities:} We use four topologies, namely \textbf{com-DBLP, com-Youtube, com-Amazon, com-Livejournal}  \cite{yang2015defining}. The Amazon network was collected by crawling Amazon.com. It is based on product co-purchasing recommendations that is, if a product i is frequently co-purchased with product j then the graph contains an undirected edge from i to j. DBLP is a co-authorship network where two authors are connected if they have published at least one paper together. In the Youtube social network, users form friendship relationships
with each other  along the lines of Facebook. Finally livejournal is an online blogging community where users can declare friendships. All these community datasets have edges close to or exceeding 1 million thus making them ideal candidates for large scale replication.
   
\end{itemize}

\subsection{Baseline and Evaluation Metrics}

We use the real world topology with our three synthetic topologies and compare them (\textbf{Figure 4}) using five well accepted characteristics which are: 
clustering coefficient distribution, node degree distribution, hop length distribution, k-core vs core node distribution, and k-core vs core edge distribution.  
We also measured the resultant global clustering coefficient of the synthetic network topologies and present those results in \textbf{Table 2}. For large datasets, that is, the last 8 rows of Table 1 we use step size varying from 1000 (email-Enron, loc-gowalla, loc-brightkite), 10,000 (com-amazon, dblp), 20,000 (com-youtube) to 40,000 (com-lj). Exact values will not affect the relative accuracy wrt to the baseline which is our measurement goal. For the rest of the dataset Step is 1, so we remove it altogether from the algorithm . 
Finally we also run the SIR model based epidemiological simulation on these datasets and their synthetic counterparts and show the infection and recovery count of the affected members of the population in \textbf{Figure 3}.  The algorithm DEG would not run in most cases as pointed out in the beginning of the algorithm section.
We use the least changed algorithm wrt DEG as our baseline, which in this case happens to be \textbf{\Baseline} and we show how  \textbf{\sage} and \textbf{\siege} outperforms it and ipsofacto outperforms DEG as well.

\begin{figure*}[t]
\centering
  \begin{subfigure}[b]{.24\linewidth}
    \centering
    \includegraphics[width=.99\textwidth, height = 4 cm ]{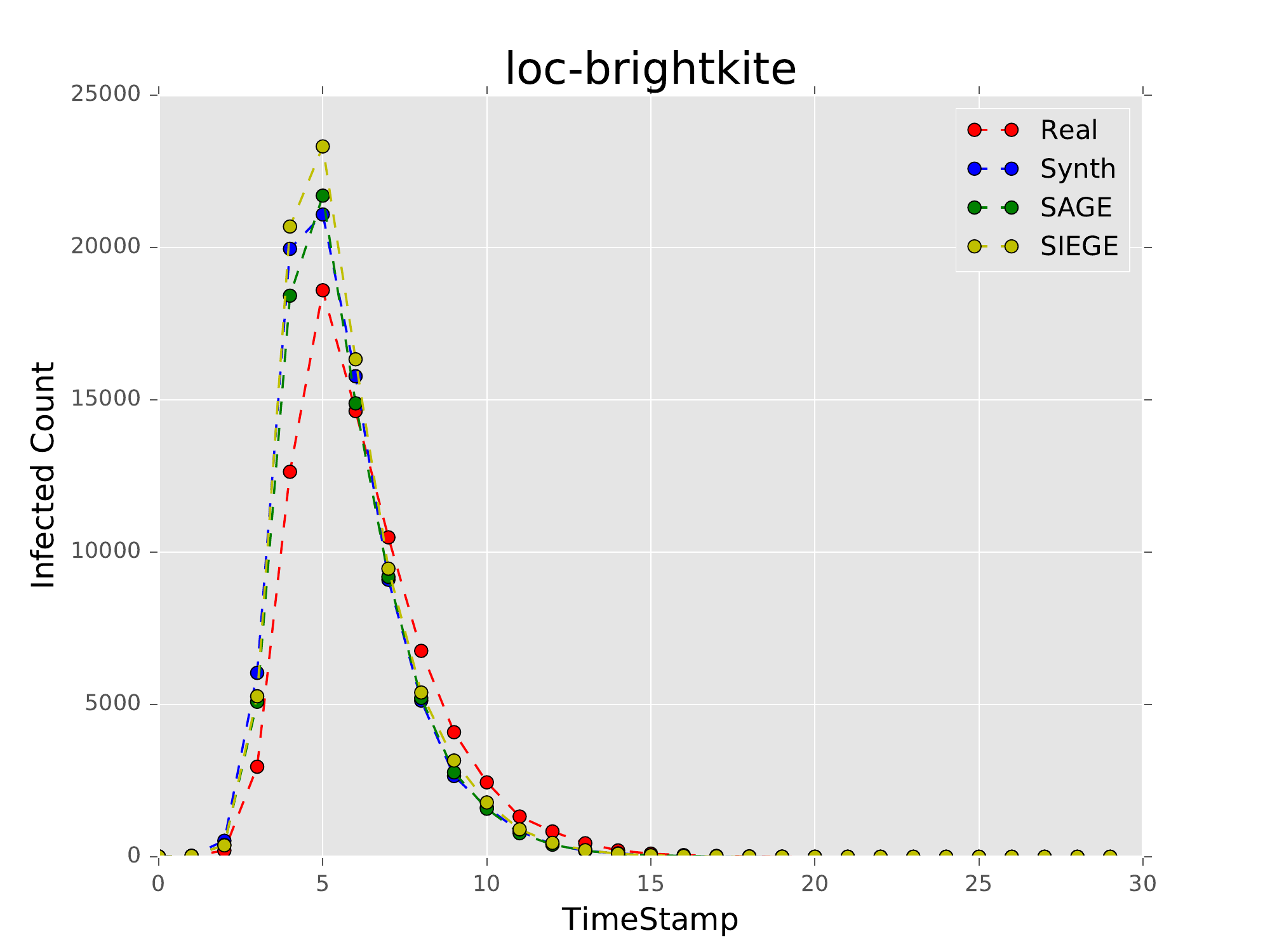}
    \caption{}\label{fig:1a}
  \end{subfigure}%
  \begin{subfigure}[b]{.24\linewidth}
    \centering
    \includegraphics[width=.99\textwidth, height = 4 cm]{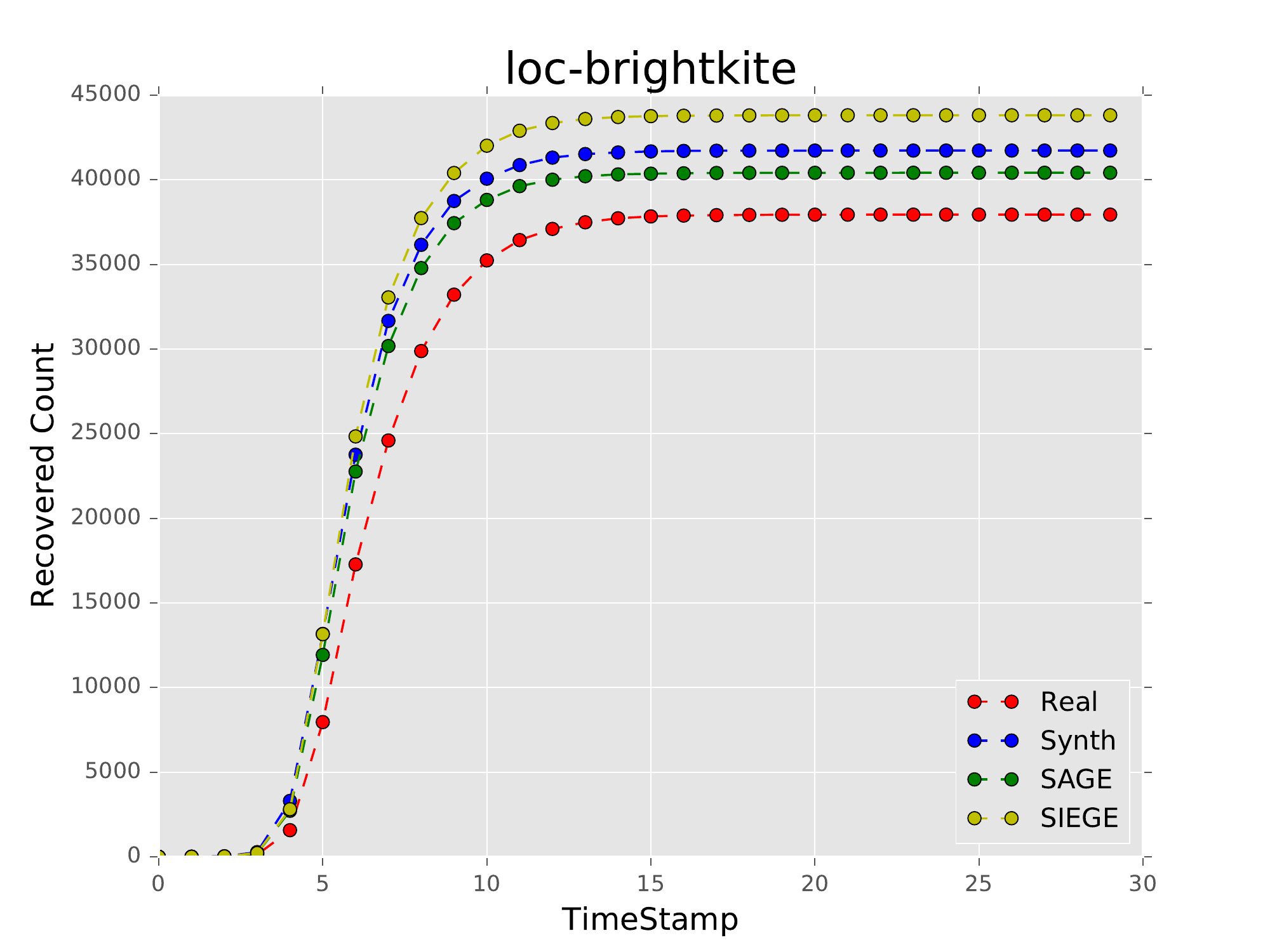}
    \caption{}\label{fig:1b}
  \end{subfigure}%
  \begin{subfigure}[b]{.24\linewidth}
    \centering
    \includegraphics[width=.99\textwidth, height = 4 cm]{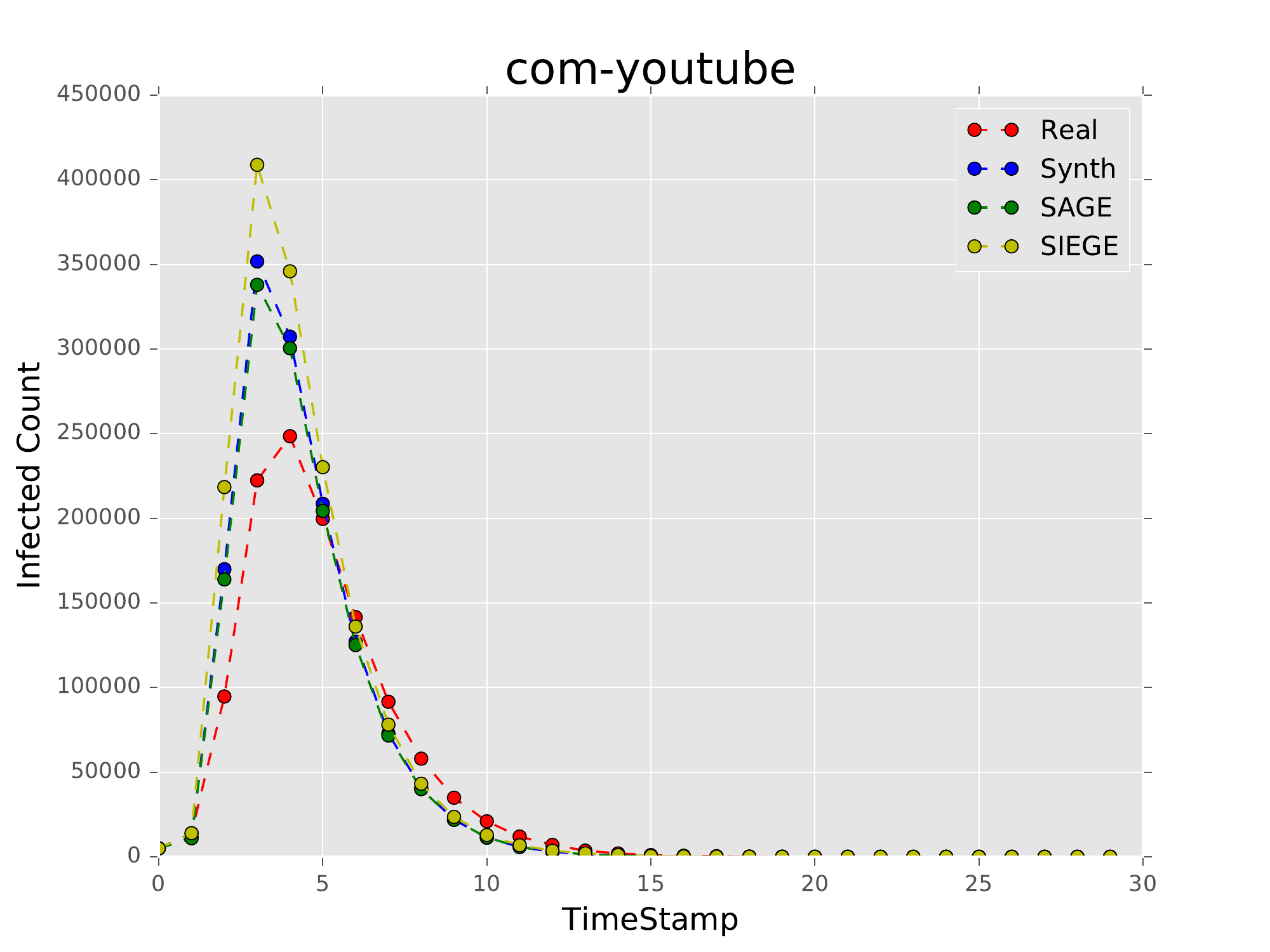}
    \caption{}\label{fig:1c}
  \end{subfigure}%
  \begin{subfigure}[b]{.24\linewidth}
    \centering
    \includegraphics[width=.99\textwidth, height = 4 cm]{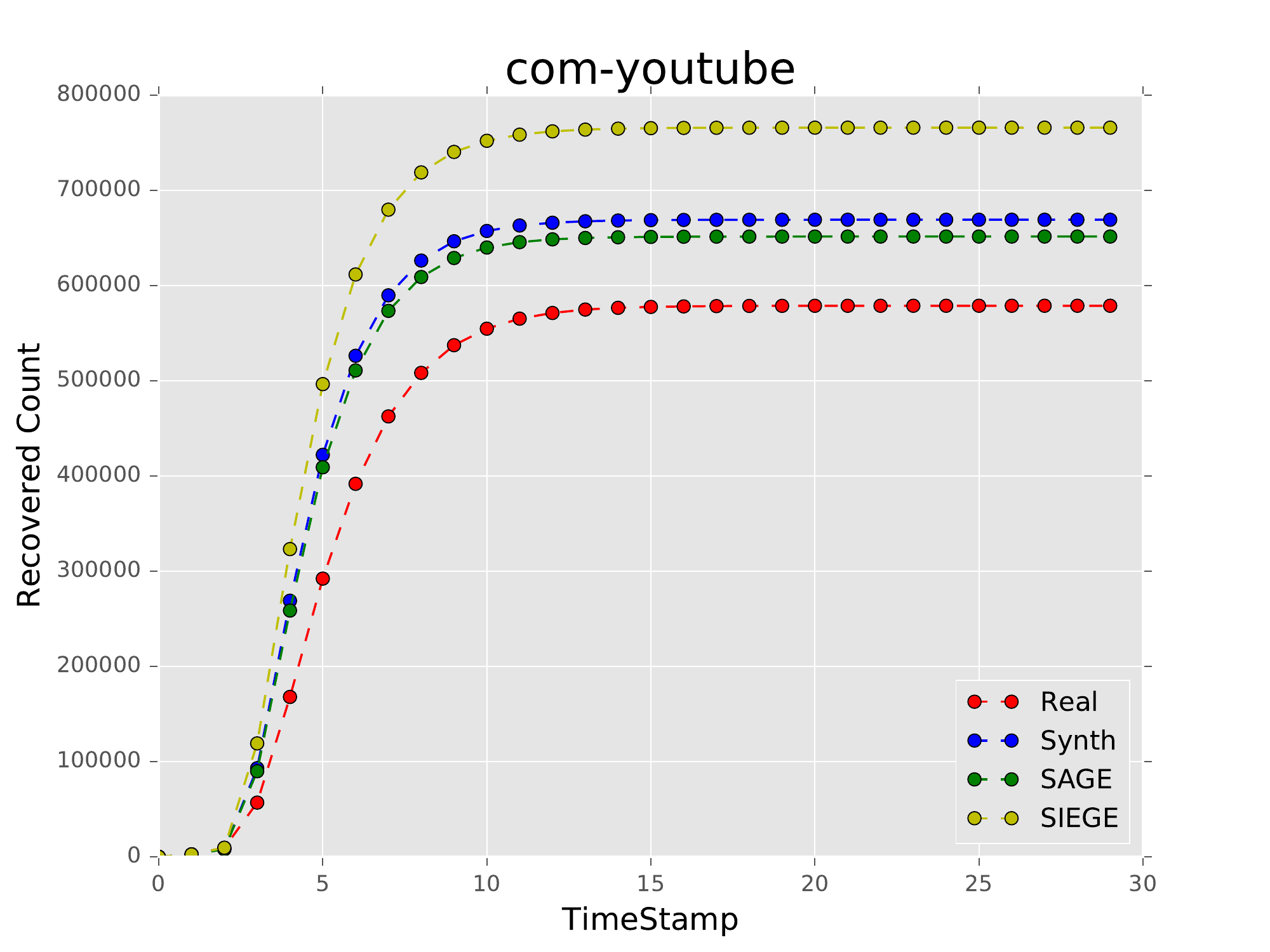}
    \caption{}\label{fig:1d}
  \end{subfigure}\\%
  \caption{(a,c) represents the infected population count of the epidemic, wherein we see the expected spike patterns are present in both of the datasets and their synthetic counterpart. Overall $\sage$ stays closest to the real world, followed by $\Baseline$ and then $\siege$. This is expected as $\siege$ has additional links created owing to its edge bridge augmentation which provides more route for the infection to spread to new nodes and so it would stray from the real world in terms of absolute count. (b,d) represents the recovered population where it can be seen that $\sage$ outperforms $\Baseline$ and $\siege$. Overall $\sage$ maintains greater fidelity in terms of reaction to SIR epidemic. }
\end{figure*}

\begin{table}[ht]
\centering
\small
\caption{$\sage$ outperforms baseline $\Baseline$ as well as $\siege$ for collaboration networks in terms of matching upto the $Real$ world topology, while for autonomous networks $\siege$ does the same. Both however, in almost all cases outperforms the baseline $\Baseline$. All of them are improvements on DEG which would give N/A in all these cases.$EB_{Count}$ captures edge bridge count}

\begin{tabular}{ |c||c|c|c|c|c|  }
 \hline
 \multicolumn{6}{|c|}{$\textrm{Global \ Clustering \ Coefficient}$} \\
 \hline
 $\textrm{Topology}$& $EB_{Count}$ & $cf_{Real}$  & $cf_{\Baseline}$&$cf_{\sagesmall}$ & $cf_{\siegesmall}$\\
 \hline 
egoFacebook   &75 &0.60554   & \textbf{0.17150} & 0.13534   & 0.13324\\
caGrQc & 1142 & 0.52960  & 0.64378   & \textbf{0.48748} & 0.37972\\
caHepTh & 2030 & 0.47143  & 0.59098 & \textbf{0.46831}	 & 0.22546\\
oregon010331  & 3799 &0.29700 & 0.82920& 0.68256  & \textbf{0.47830}\\
oregon010407 & 3848 &0.29210  & 0.64214 & 0.68730 & \textbf{0.47380}\\
oregon010414 & 3853 &0.29540  & 0.82926   & 0.67889 & \textbf{0.47723}\\
oregon010421 & 3855 &0.29680 &  0.82903 & 0.68167 & \textbf{0.47891} \\
oregon010428 & 3844 &0.29400 &  0.82948 & 0.68310 & \textbf{0.47353} \\
oregon020331 & 3274&0.50090 &  0.79330	&0.63260	&\textbf{0.49710} \\
oregon020407 & 3332&0.34630 &  0.79262	&0.63577	&\textbf{0.49019} \\
oregon020414 & 3316&0.34730 &  0.73276	&0.63130	&\textbf{0.49460} \\
oregon020421 & 3294&0.34960 &  0.79208	&0.63152	&\textbf{0.49562} \\
oregon020428 & 3283&0.34720 &  0.79165	&0.63136	&\textbf{0.49750} \\
oregon020505 & 3282&0.34610 &  0.79305	&0.63379	&\textbf{0.49737} \\
oregon020512 & 3296&0.34650 &  0.79243	&0.63581	&\textbf{0.50218} \\
oregon020519 & 3320&0.34840 &  0.79332	&0.63709	&\textbf{0.49836} \\
oregon020526 & 3351& 0.41430&  0.78936	&0.63203	&\textbf{0.49721} \\
email-Enron & 10714&0.49700 &  0.54747	&\textbf{0.44578}	&0.36447 \\
loc-gowalla & 54351&0.23670 &  0.51262	&0.40931	&\textbf{0.33317} \\
loc-brightkite & 23155&0.17230 &  0.59817	&0.47045	&\textbf{0.33137} \\
com-amazon & 312191&0.39670 &  0.42287	& \textbf{0.40765}	&0.36733 \\
com-dblp & 47103&0.63240 & \textbf{0.48473}	&0.42230	&0.36637 \\
com-youtube & 667090 & 0.08080 & 0.73160 &0.58609	&\textbf{0.33239} \\
as-skitter & 232141&0.25810 & 0.43069	&0.34795	&\textbf{0.32145}\\
com-liveJournal &821887 & 0.28427& 0.38987& \textbf{0.28572}& 0.25469 \\
 \hline
\end{tabular}
\end{table}

\section{Results and Discussions}

For want of space all the graphical plots cannot be shown. We present the results for \textbf{as-skitter, com-LiveJournal} and 
\textbf{oregon010428} primarily.

\begin{itemize}[leftmargin = 0.4cm]

\item \textbf{How similar are triadic closures at the node level?}

\begin{itemize}[leftmargin = 0.4cm]
\item \textbf{Clustering Coefficient Distribution:} 
As can be seen in \textbf{Figure 4 (a,b,c)}, the clustering coefficient distribution for \textbf{\Baseline, \sage and \siege } follows the trend of the real world. The variance as the node degree increases is comparatively lesser for the three synthetic topologies as opposed to the real world network.
The global clustering coefficient values presented in Table 2, that is, $CF_{\Baseline}$, $CF_{\sage}$, $CF_{\siege}$ consistently showed a decrement in value for any given dataset. This is a direct consequence of the fact that \textbf{\sage} and \textbf{\siege} allows for more single edge creation that increases the total number of triples in the topology. 
In terms of actual values $CF_{\sage}$ was closest to the real world collaboration topologies \textbf{ca-Gr-Qc, com-amazon, com-liveJournal} and \textbf{ca-Hep-Th}. For the autonomous systems based networks and location based OSN \textbf{\siege} clearly performed best. 
\end{itemize}


 
\item \textbf{Can the models reproduce the reachability nature of the real world data?}
\begin{itemize}[leftmargin = 0.4cm]
\item \textbf{Shortest Path Distribution:}
As can be seen in \textbf{Figure 4 (d,e,f)} the Synthetic topologies matches quite well with the real world topology. This was observed across all the datasets, except for the last seven large datasets in table 1 for which the all pair shortest path computation time bloated out and hence we cannot comment on them. So we show the results for oregon010331, oregon010414 in figure 4 (d,f) instead.
The distributions are either unimodal or vaguely bimodal for the real world datasets. The synthetic topologies are unimodal. However, as stated earlier they still capture the trend.
\end{itemize}

\item \textbf{What about other clustering heuristics?}

\begin{itemize}[leftmargin = 0.4cm]
\item \textbf{k-core vs core node distribution:} 
The real world topologies tend to have more nodes participating in k-core when k is less. As k increases the total number of nodes participating tends to decrease. For all the twenty-five real world datasets,(A sample of which has been provided in \textbf{Figure 4 (g,h,i)}) this property was seen and the synthetic topologies captured this characteristics as well.
Among the three synthetic generation algorithm \textbf{\siege} is performing better in general.
\item \textbf{k-core vs core edge distribution:}
The k-core vs core participating edges distribution also matched well for all the datasets. The broad trend of more participating edges in less valued core was captured effectively by all the algorithms as evident in \textbf{Figure 4 (j,k,l)}.
\end{itemize}
\item \textbf{Is there a power law?} 

\begin{itemize}[leftmargin = 0.4cm]
\item \textbf{Node Degree Distribution} 
As shown in \textbf{Figure 4 (m,n,o)} the heavy tail power law distribution is maintained making these algorithms suitable for generating synthetic topologies imitating real world networks. 
A wide variance is observed even on a log-log scale at node degree axis for \textbf{\Baseline} which is our baseline, which stabilizes as the heavy tail is approached. This is not the case for \textbf{\sage} and \textbf{\siege}. This trend seems to be consistent across all the datasets. \siege clearly outperforms \Baseline in all these datasets and captures the heavy tail very effectively as the node degree increases.
\end{itemize}

\item \textbf{Does the model shed light into the behavioral nature of 
graphs?}

\begin{itemize}
\item \textbf{SIR Epidemiological Simulation Outcome}
We choose the disease transmission rate to be 0.3 and the recovery rate to be 0.5 for the population. As shown in \textbf{Figure 3(a,c)} this results in a spike in the total infected count before the epidemics starts dying out. \textbf{\sage} stays closest to the real world followed by \textbf{\Baseline} followed by \textbf{\siege}. 
\textbf{Figure 3(b,d)} represents the recovered population count with \textbf{\sage} performing better than the rest. 
\end{itemize}

\item \textbf{What do we lose as we scale up?}
\begin{itemize}
\item Experimental evidence shows us we do not lose much in terms of conformity to the real world characteristics across all the datasets, except 
for the case of
absolute values of the average clustering coefficient distribution. The global value maintains considerable fidelity as shown in Table 2. In this paper we are not specifically doing any study of the effect of variations of step size given any single dataset. Intuitively we expect accuracy to fall as step size increases.   
\end{itemize}
\end{itemize}
\section{Conclusion and Future work} 
Our work proposes stable algorithms for synthetic network generation at topological level. They are demonstrably better than current state-of-the-art
algorithms. As our future work we plan to incorporate attribute values for the nodes as well as principles of homophily and diversity into the topology generation process. We hope that would create a richer synthetic dataset for the research community.

\begin{figure*}[t]
\centering       
  \begin{subfigure}[b]{.24\linewidth}
    \centering
    \includegraphics[width=.99\textwidth, height = 4 cm]{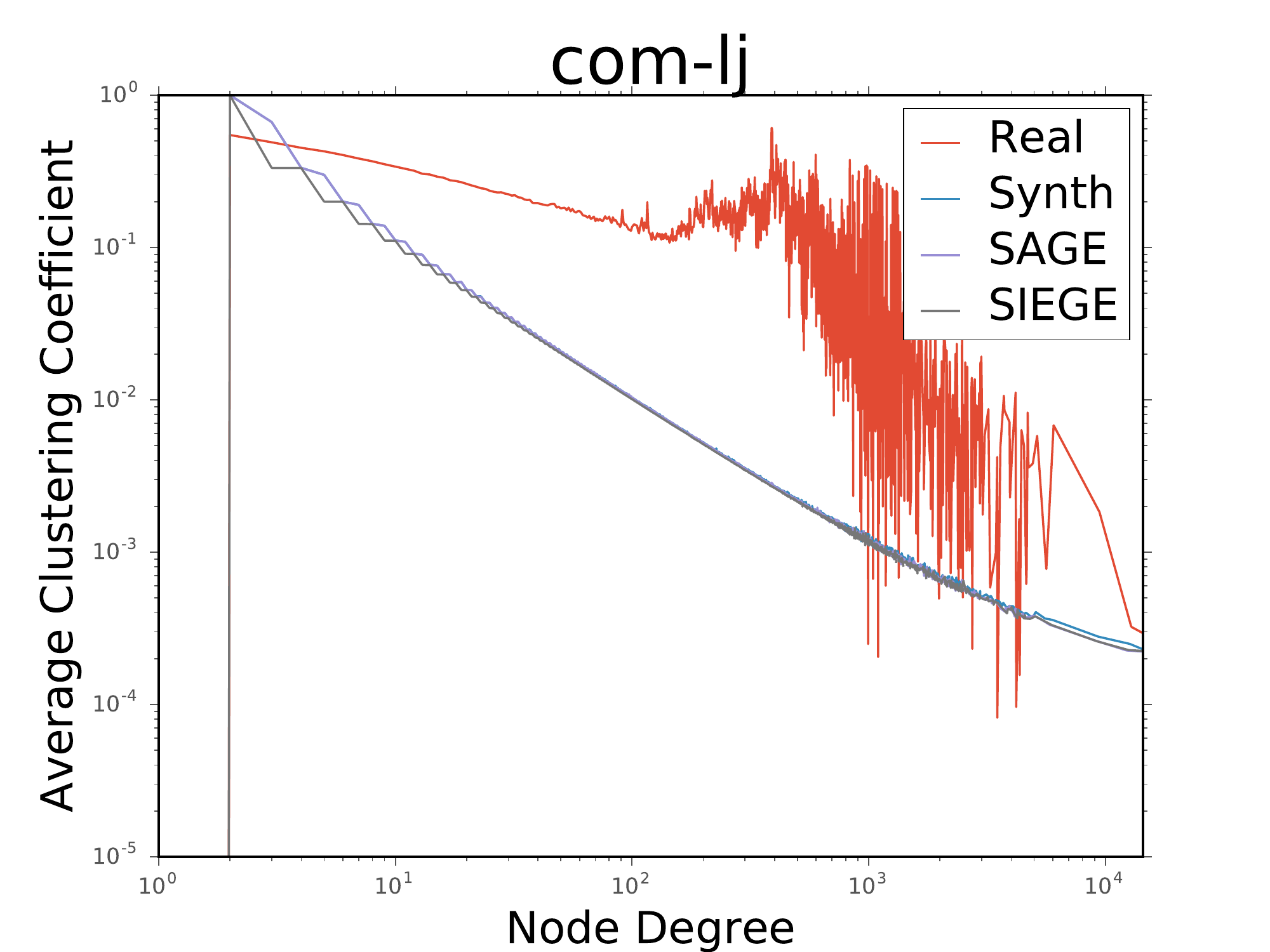}
    \caption{}\label{fig:1a}
  \end{subfigure}%
  \begin{subfigure}[b]{.24\linewidth}
    \centering
    \includegraphics[width=.99\textwidth, height = 4 cm]{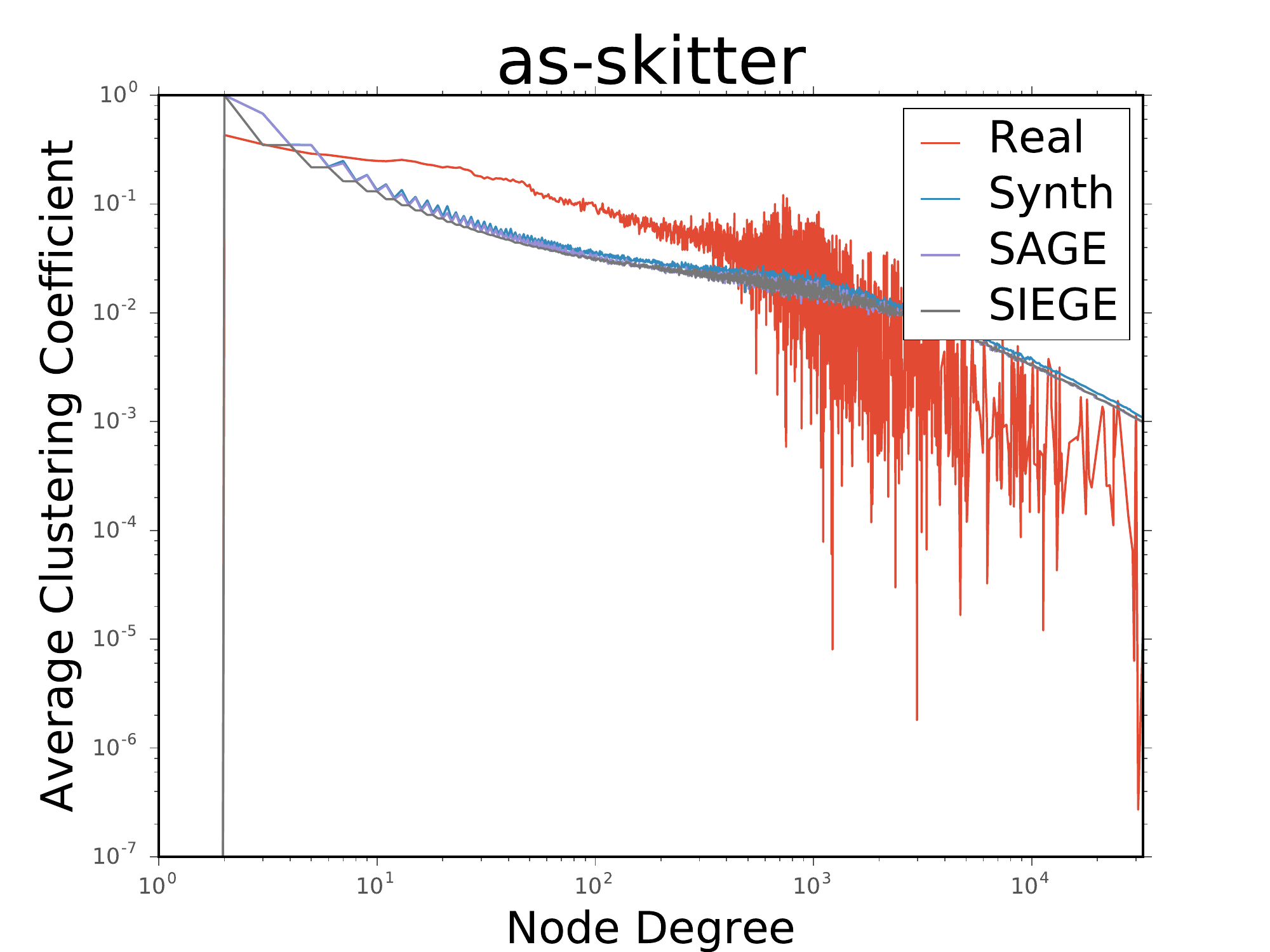}
    \caption{}\label{fig:1b}
  \end{subfigure}%
  \begin{subfigure}[b]{.24\linewidth}
    \centering
    \includegraphics[width=.99\textwidth, height = 4 cm]{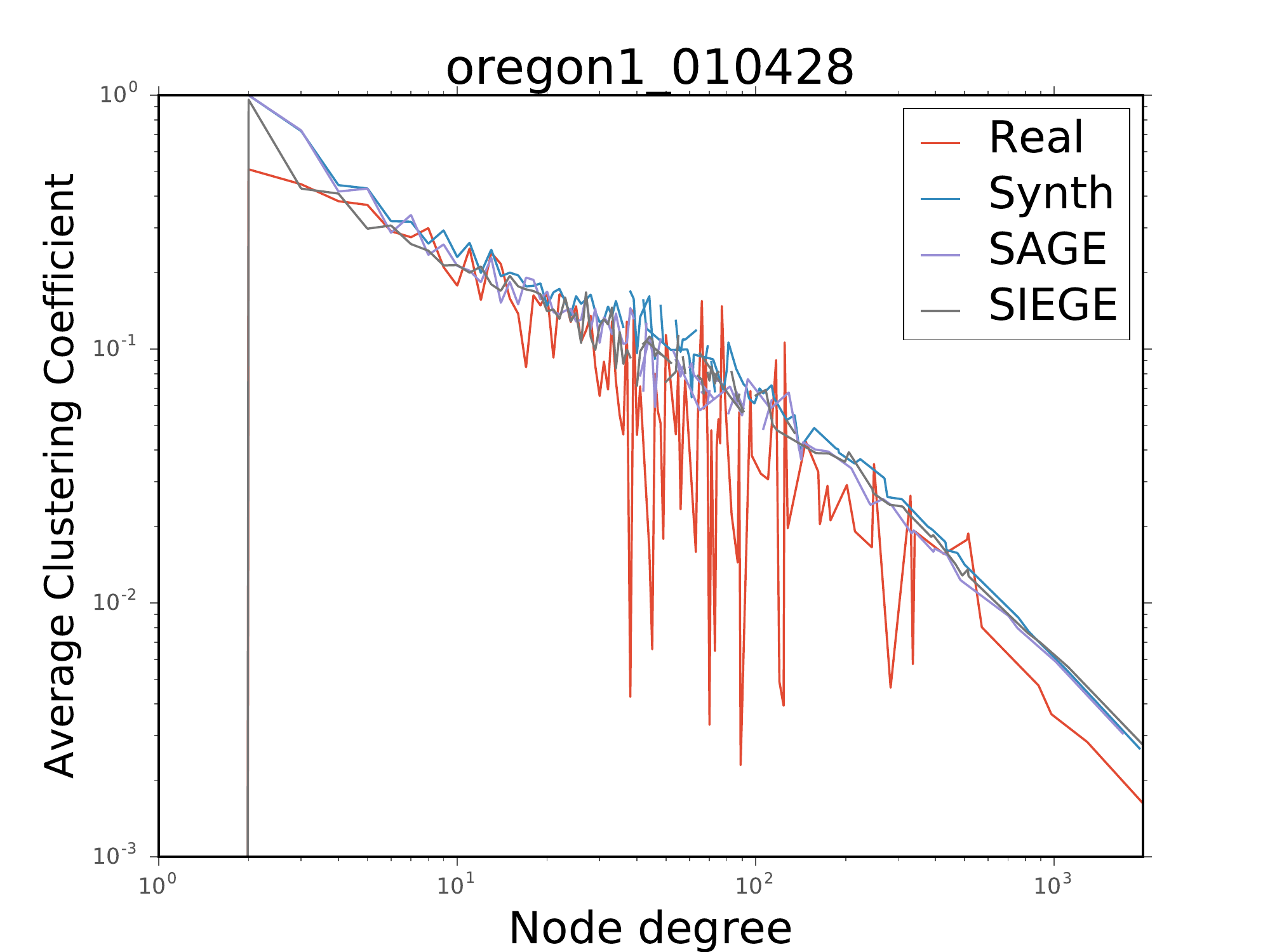}
    \caption{}\label{fig:1c}
  \end{subfigure}%
  \begin{subfigure}[b]{.24\linewidth}
    \centering
    \includegraphics[width=.99\textwidth, height = 4 cm]{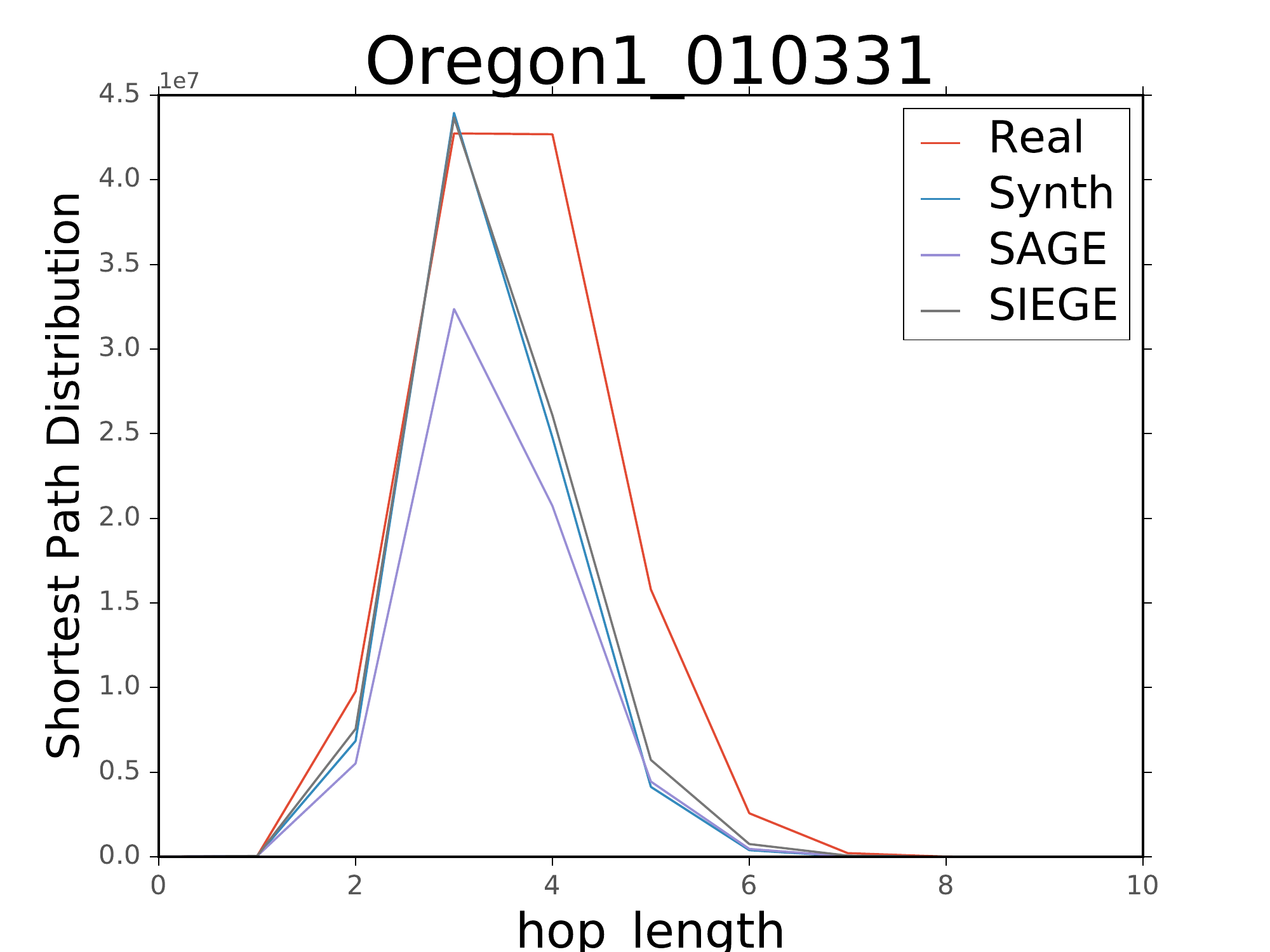}
    \caption{}\label{fig:1d}
  \end{subfigure}\\%
  \begin{subfigure}[b]{.24\linewidth}
    \centering
    \includegraphics[width=.99\textwidth, height = 4 cm]{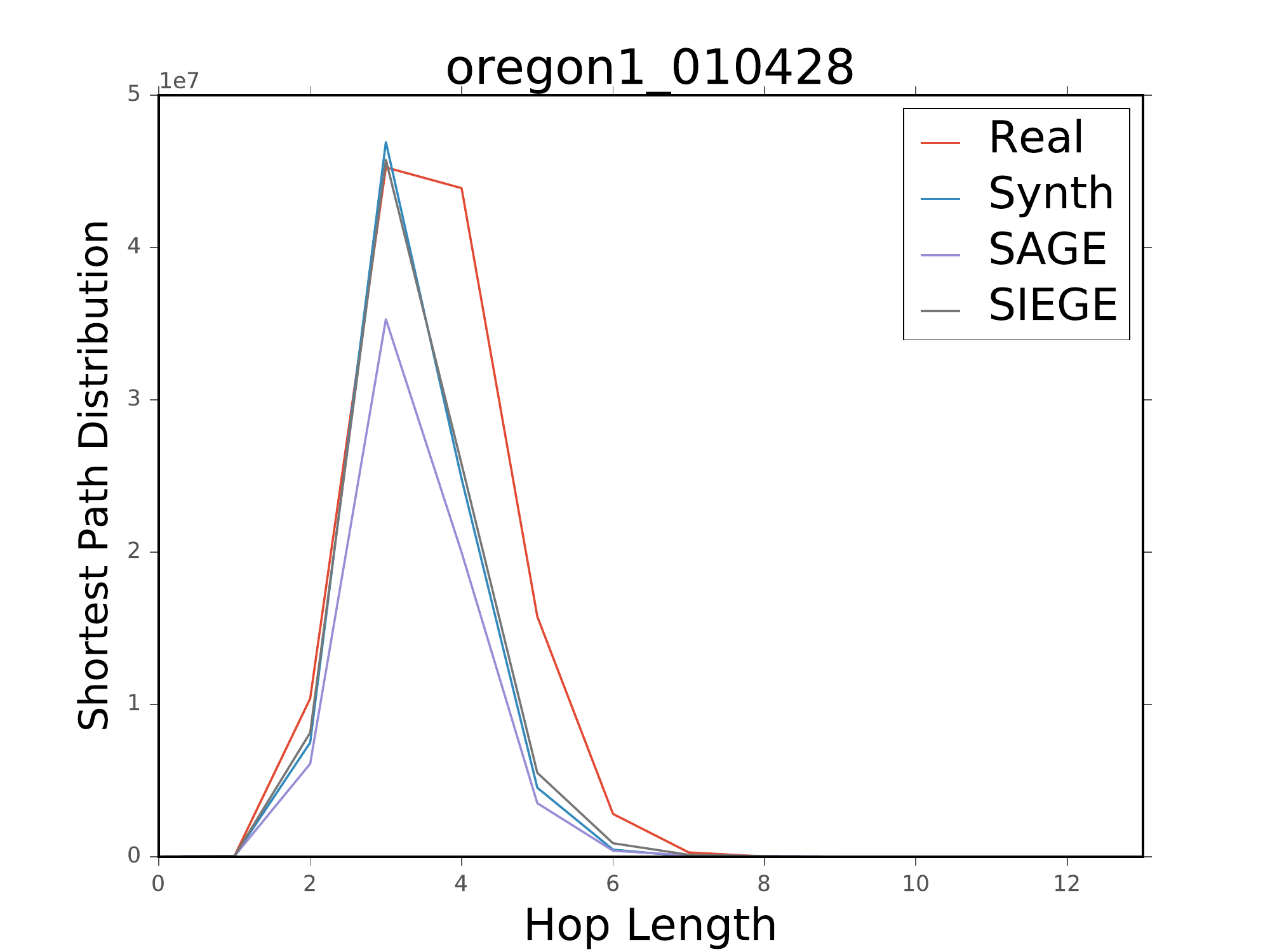}
    \caption{}\label{fig:1e}
  \end{subfigure}%
  \begin{subfigure}[b]{.24\linewidth}
    \centering
    \includegraphics[width=.99\textwidth, height = 4 cm]{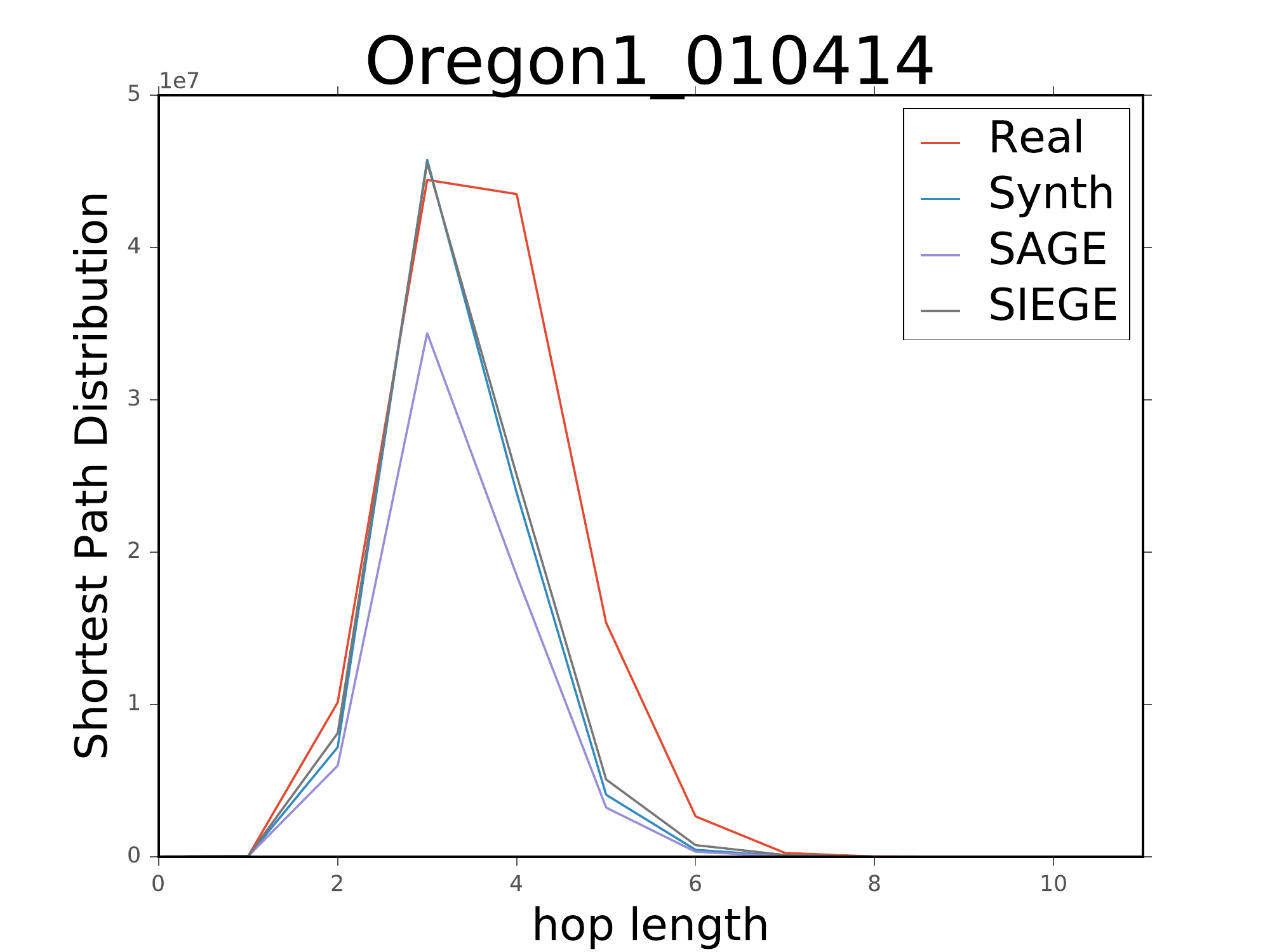}
    \caption{}\label{fig:1f}
  \end{subfigure}%
  \begin{subfigure}[b]{.24\linewidth}
    \centering
    \includegraphics[width=.99\textwidth, height = 4 cm]{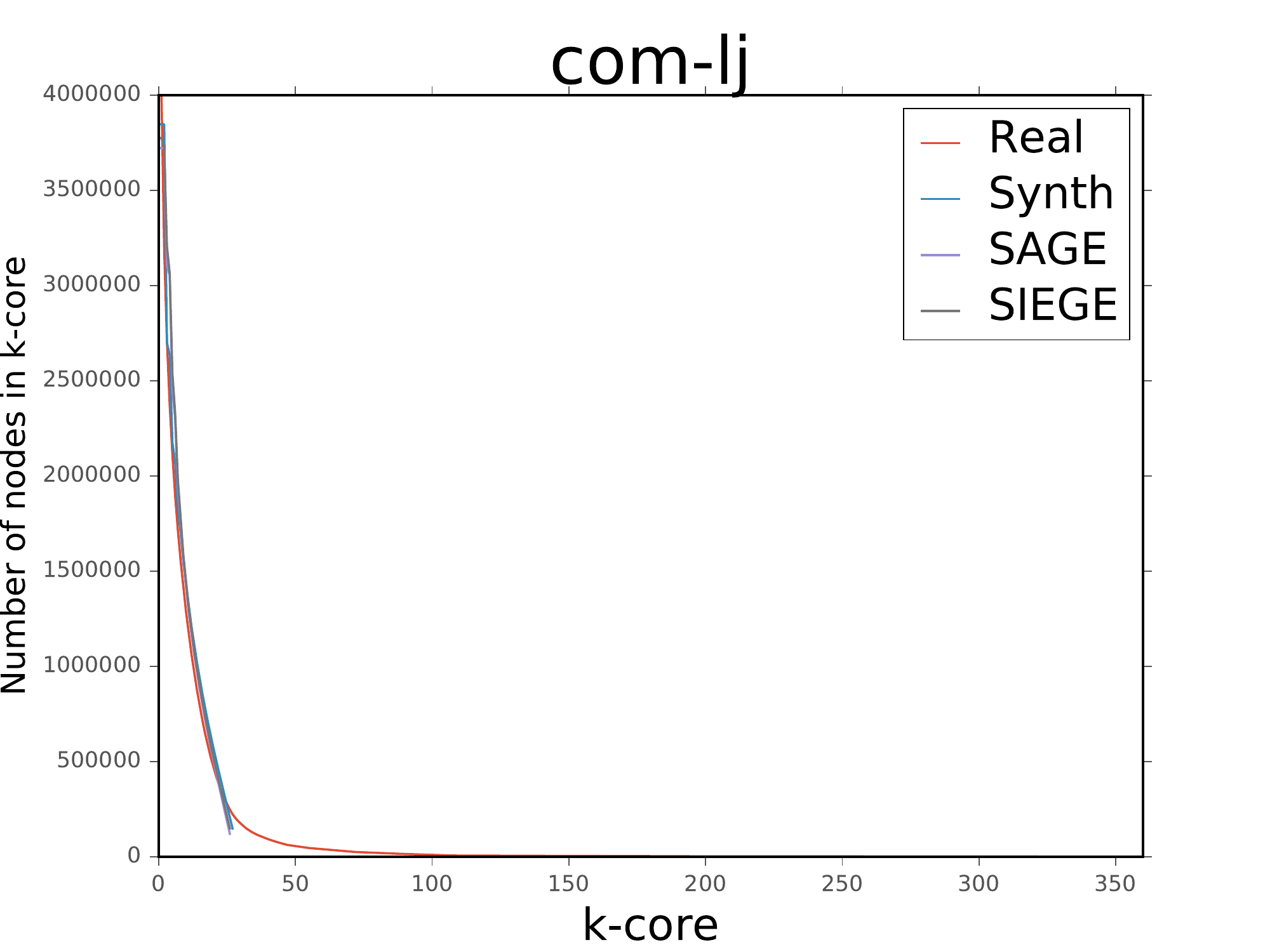}
    \caption{}\label{fig:1g}
  \end{subfigure}%
  \begin{subfigure}[b]{.24\linewidth}
    \centering
    \includegraphics[width=.99\textwidth, height = 4 cm]{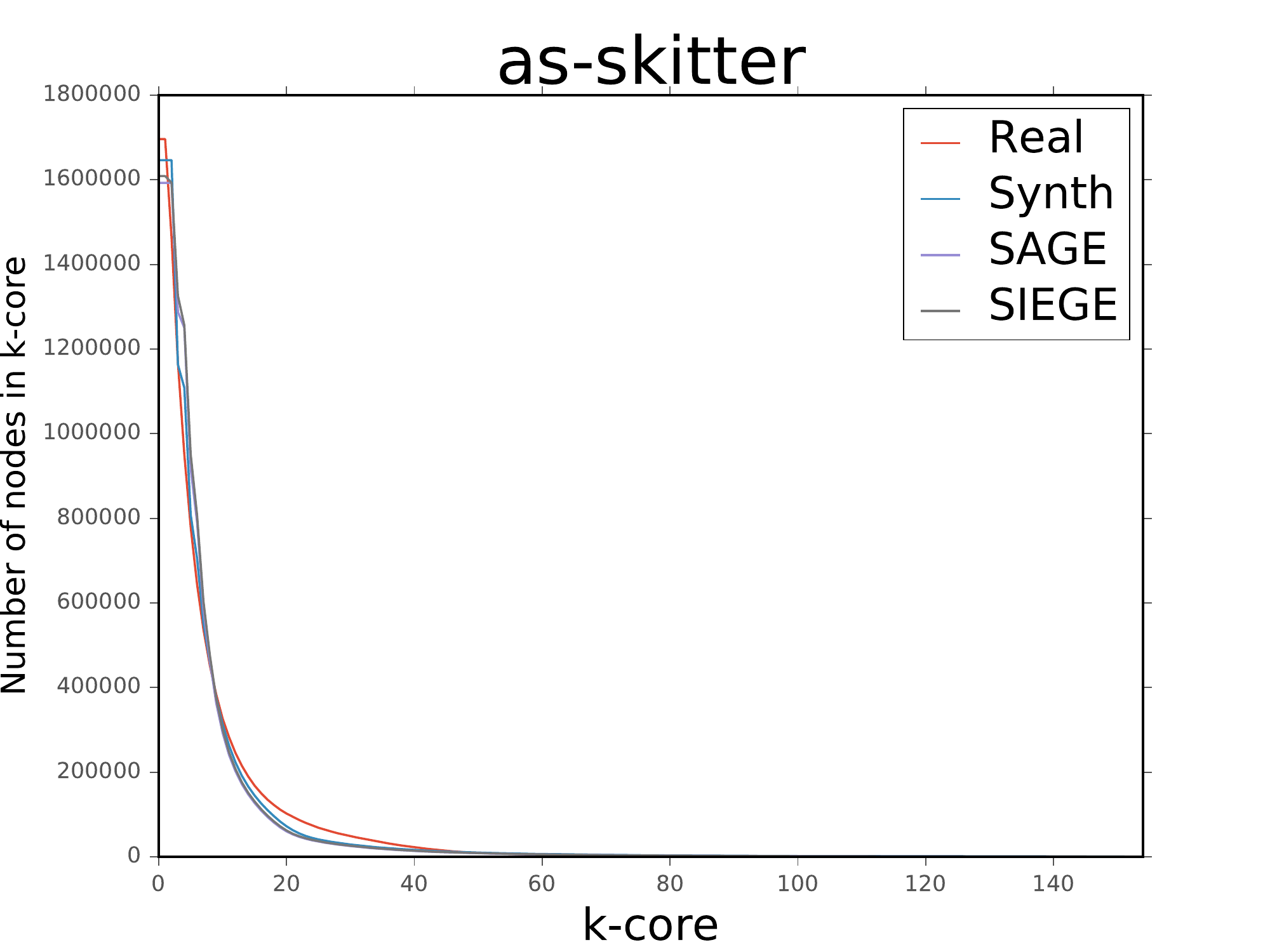}
    \caption{}\label{fig:1h}
  \end{subfigure}\\%
  \begin{subfigure}[b]{.24\linewidth}
    \centering
    \includegraphics[width=.99\textwidth, height = 4 cm]{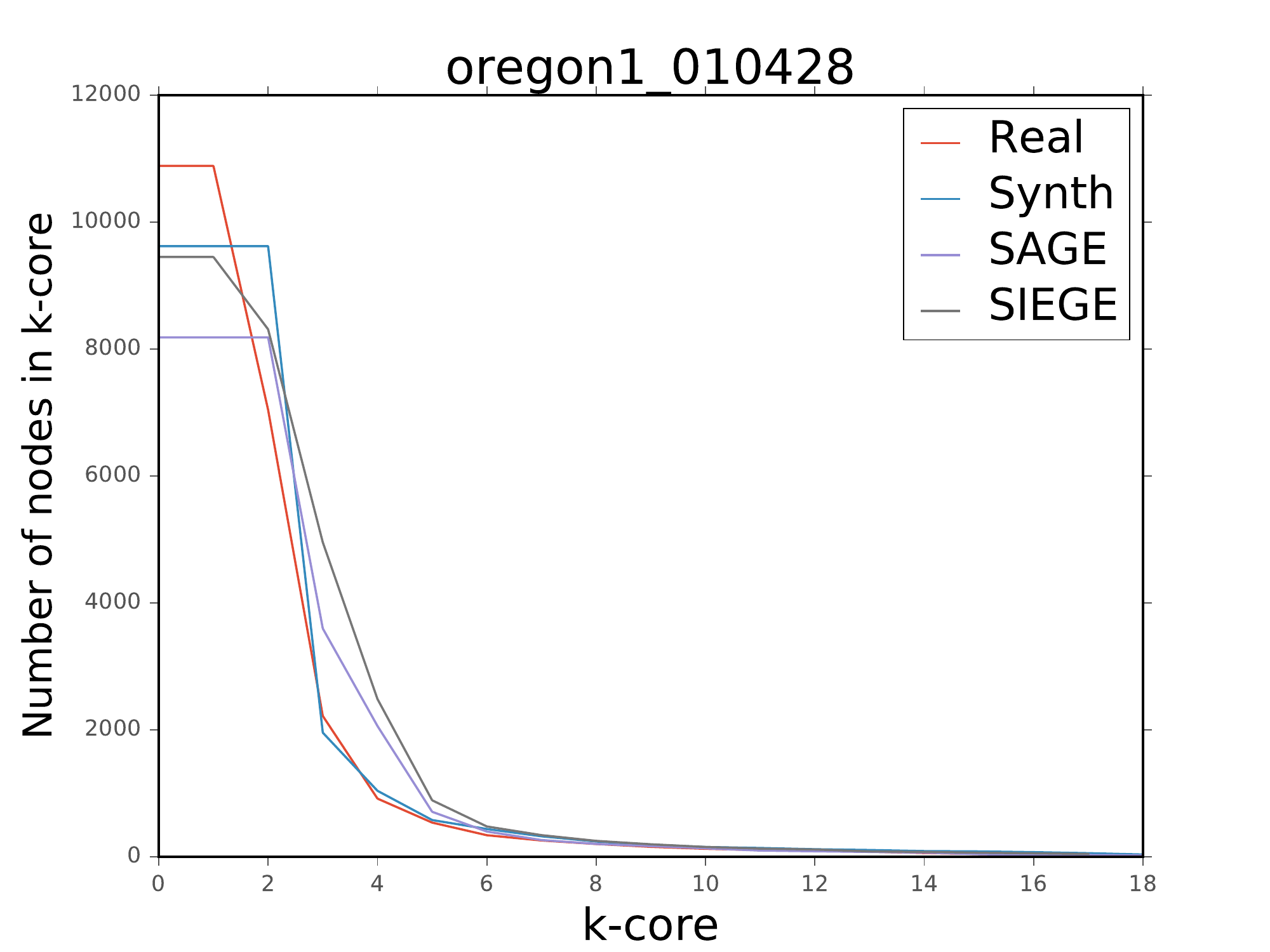}
    \caption{}\label{fig:1i}
  \end{subfigure}%
  \begin{subfigure}[b]{.24\linewidth}
    \centering
    \includegraphics[width=.99\textwidth, height = 4 cm]{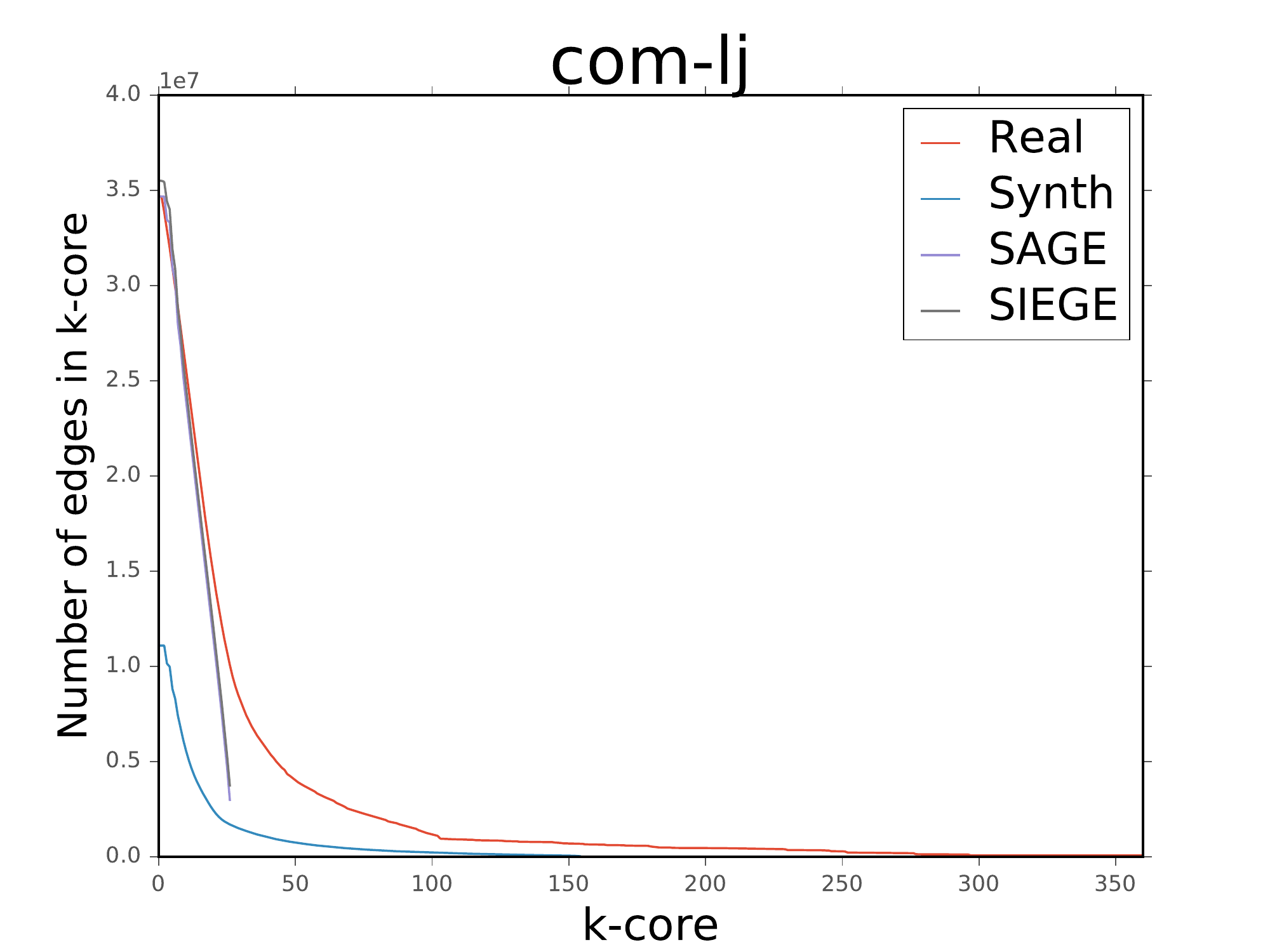}
    \caption{}\label{fig:1j}
  \end{subfigure}%
  \begin{subfigure}[b]{.24\linewidth}
    \centering
    \includegraphics[width=.99\textwidth, height = 4 cm]{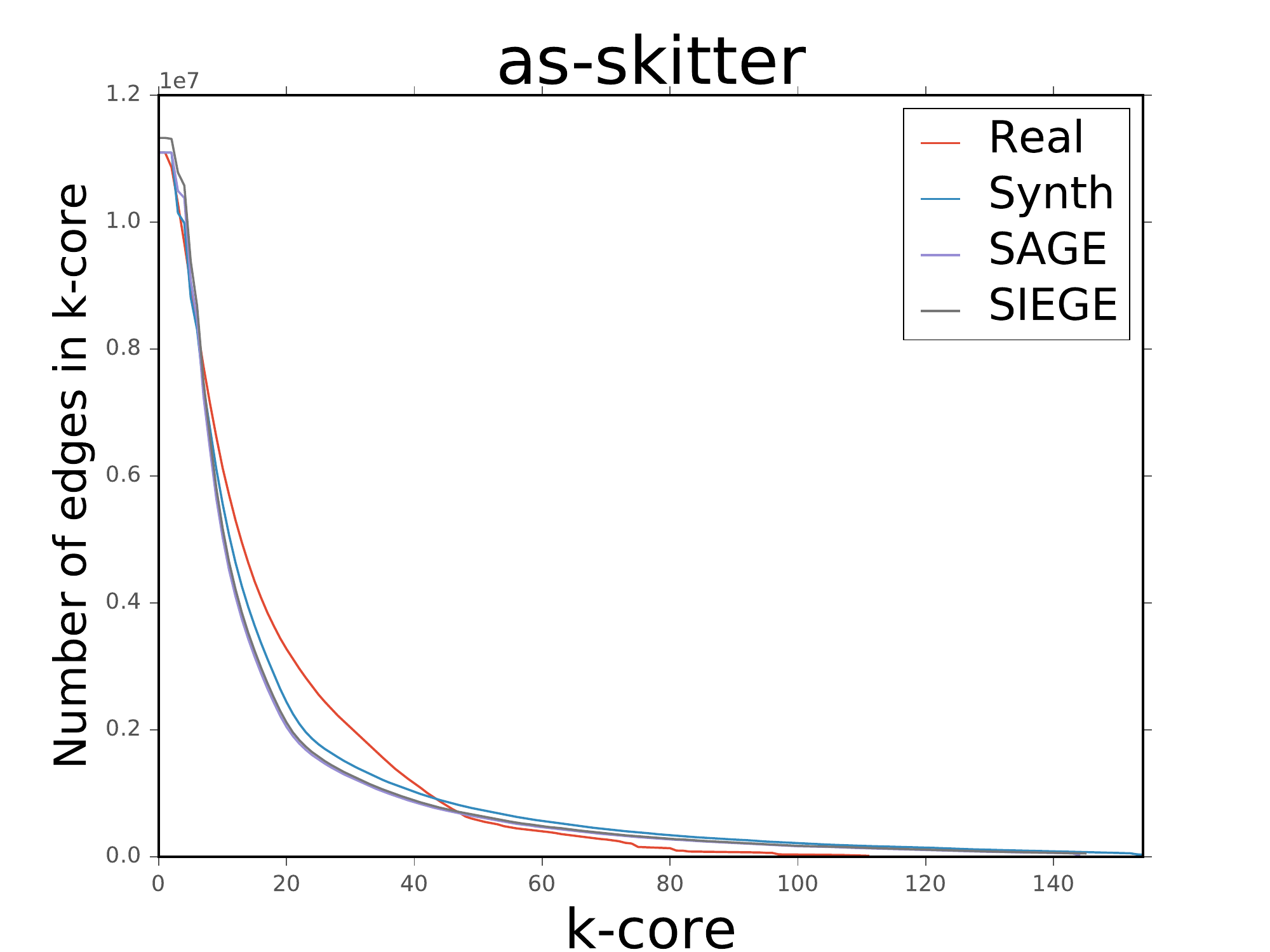}
    \caption{}\label{fig:1k}
  \end{subfigure}%
  \begin{subfigure}[b]{.24\linewidth}
    \centering
    \includegraphics[width=.99\textwidth, height = 4 cm]{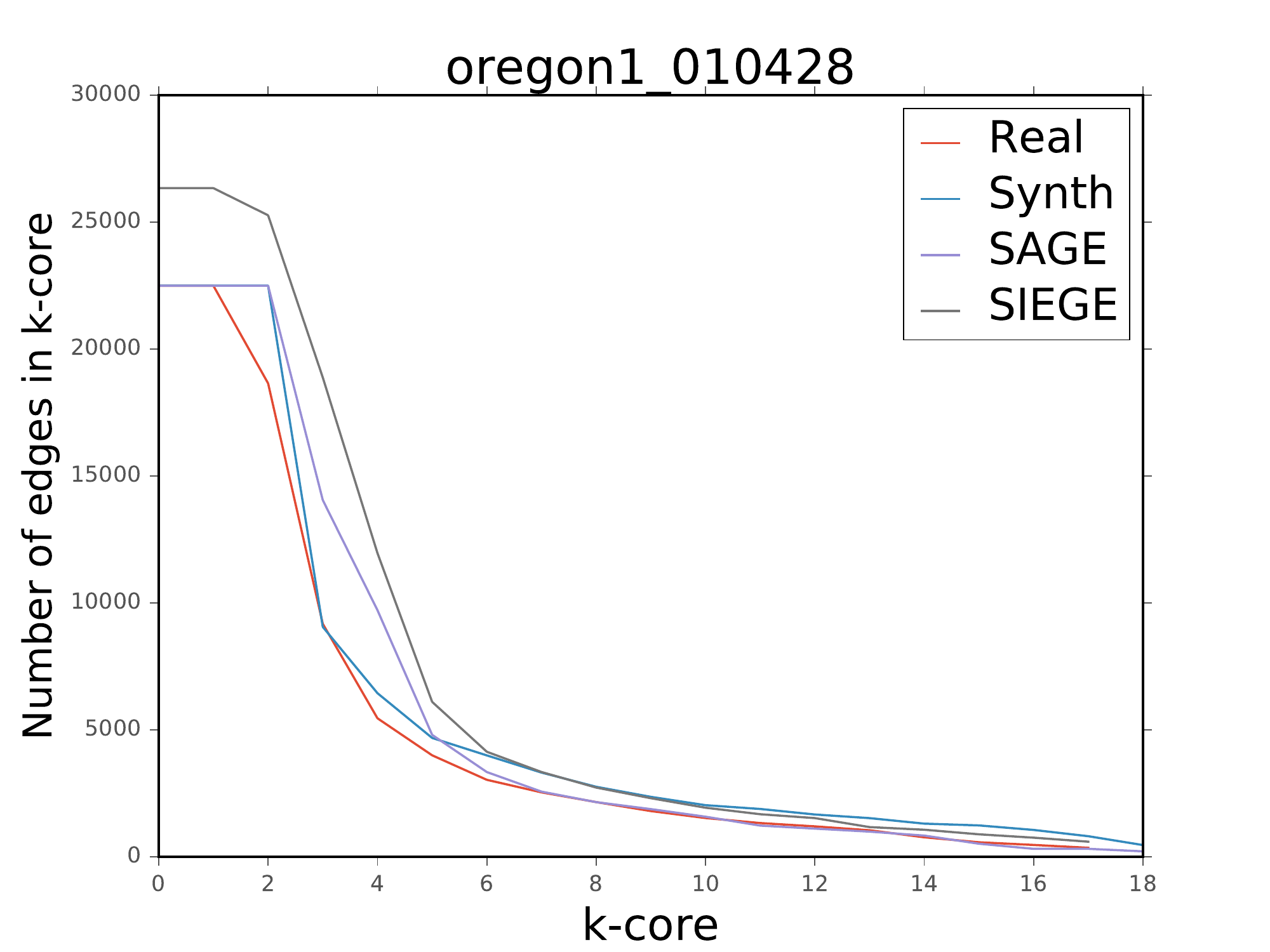}
    \caption{}\label{fig:1l}
  \end{subfigure}\\%
  \begin{subfigure}[b]{.33\linewidth}
    \centering
    \includegraphics[width=.99\textwidth]{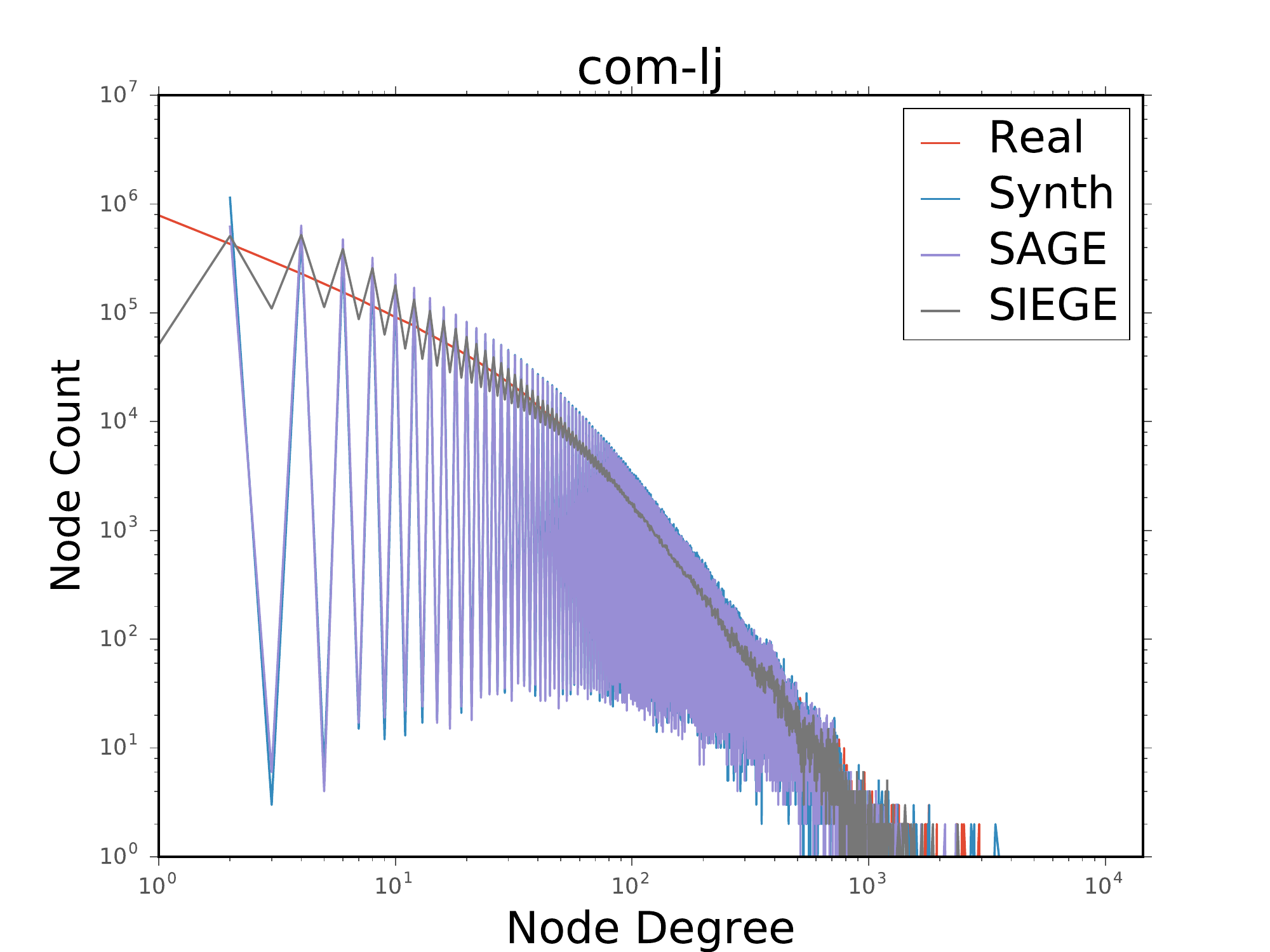}
    \caption{}\label{fig:1m}
  \end{subfigure}%
  \begin{subfigure}[b]{.33\linewidth}
    \centering
    \includegraphics[width=.99\textwidth]{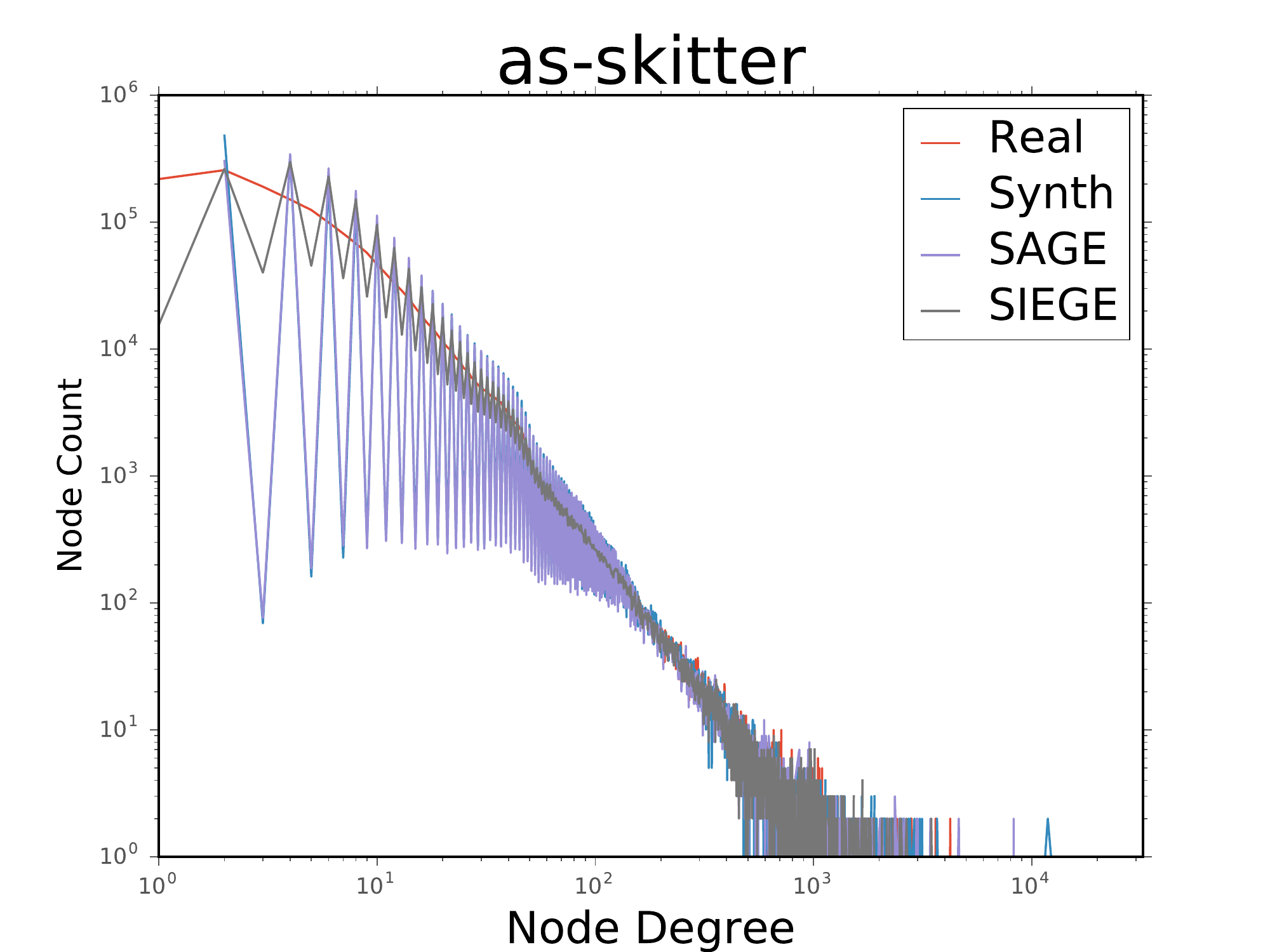}
    \caption{}\label{fig:1n}
  \end{subfigure}%
  \begin{subfigure}[b]{.33\linewidth}
    \centering
    \includegraphics[width=.99\textwidth]{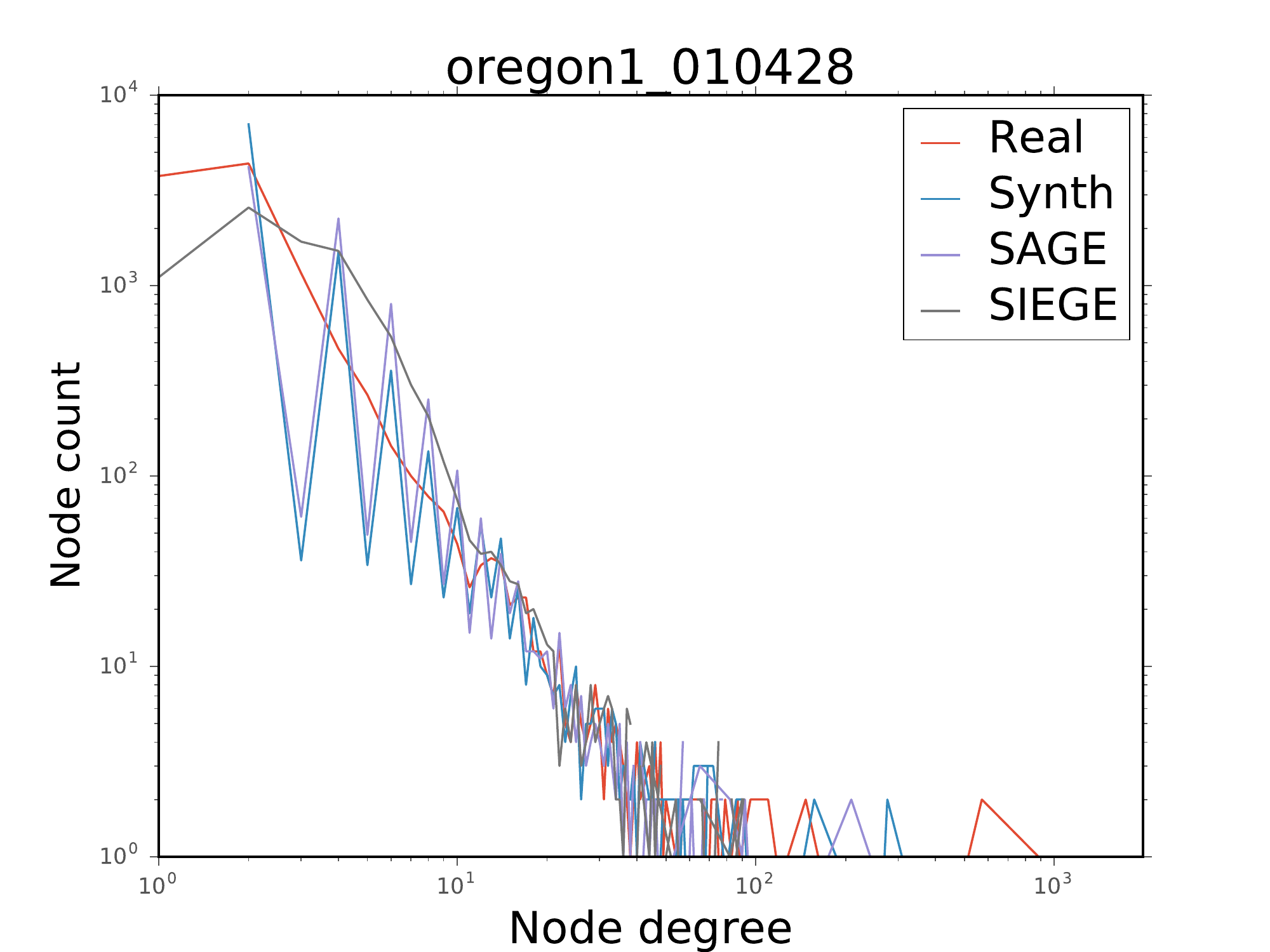}
    \caption{}\label{fig:1o}
  \end{subfigure}%
  \caption{\textbf{(a,b,c)} represents the Local clustering coefficient distribution for dataset com-lj, as-skitter and Oregon010428 respectively, on a log-log scale. We can see the real network trend is broadly captured by all 3 models. \textbf{(d,e,f)} represents the hop-length distribution, which is either unimodal or bimodal for the real world. In all the datasets this characteristic is 
mimiced very well by our models. \textbf{(g,h,i)} represents k-core vs number of nodes participating in the core and \textbf{(j,k,l)} represents k-core vs number of edges participating in the core. In all six of them our models show encouraging stability. \textbf{(m,n and o)} represents the node degree distribution wherein our proposed model \siege outperforms the baseline \Baseline in all the cases. It is noted that in all these datasets the models produce stable results as opposed to the model DEG which would stall, making the outcomes in most of these scenarios null. Thus there is a significant improvement in stability from our approach.}\label{fig:1}
\end{figure*}

\bibliographystyle{ACM-Reference-Format}

\bibliography{cf} 

\end{document}